\documentclass[aps,prl,twocolumn,showpacs,superscriptaddress,groupedaddress]{revtex4-2}  
\usepackage{graphicx}
\usepackage{dcolumn}
\usepackage{bm}

\usepackage[utf8]{inputenc}
\usepackage[T1]{fontenc}
\usepackage{mathptmx}
\usepackage{epsfig}
\usepackage{subfig,float}
\usepackage{graphicx}
\usepackage{caption}
\usepackage{xcolor}
\usepackage{float}

\usepackage{mathptmx}      
\usepackage{hyperref}
\usepackage{textcomp}
\usepackage[belowskip=-8pt,aboveskip=6pt]{caption}
\hypersetup{
    colorlinks=true,
    linkcolor=blue,
    citecolor=blue,
    filecolor=magenta,      
    urlcolor=cyan,
}

\begin{document}
	

\title{Wide elastic wave bandgap metamaterial with single phase constituent}

\author{Nitish Kumar}
\author{Siladitya Pal}

\address{Multiscale Mechanics and Multiphysics Laboratory, Mechanical and Industrial Engineering Department, Indian Institute of Technology Roorkee, Roorkee, Uttarakhand 247667, India}
%


\date{\today}

\begin{abstract}
	Accomplishing a wide elastic wave bandgap with single phase constituent is of primary interest in developing phononic metamaterials. In the present article, exploiting spatial periodicity, a single phase lattice is configured towards achieving a large frequency bandgap in sonic range. Numerical simulations reveal the presence of a comprehensive bandgap of 18 kHz in the 2 to 22 kHz range with systematically localizing the same constituent material in the lattice. Bloch wave
	modes unravel the involvement of dipole, monopole, and quadrupole resonances for wide and connected bandgaps. The existence of salient bandgaps is experimentally validated by analyzing the mechanical wave transmission.
\end{abstract}

\maketitle

Phononic metamaterials exhibit unconventional and unusual approaches to control the sub-wavelength elastic/acoustic waves~\citep{ma2016acoustic,lee2017acoustic}. Among various properties, the propagation of elastic waves is restricted over a frequency range known as bandgap. Due to the presence of spatial periodicity with multiple constituents in the structure, either local resonance~\citep{liu2000locally,ma2016acoustic} or Bragg scattering~\citep{kittel1976introduction} are manifested
leading to the formation of the bandgap. Thus, phononic metamaterials find futuristic applications such as vibration isolation~\citep{koh2014phononic,hu2017acoustic}, noise insulators/absorbers~\citep{wang2013two,chiang2011multilayered}, negative refraction~\citep{lu2007negative,oh2017doubly}, acoustic cloaking~\citep{cummer2007one,zigoneanu2014three}, and zero index medium~\citep{nguyen2010total,jing2012numerical}.
Particularly, locally resonant metamaterials consist of low frequency bandgap due to negative effective mass at dipole resonance~\citep{liu2000locally,yang2008membrane,lee2009acoustic}, and negative effective moduli near quadrupole/monopole resonances~\citep{russell1999acoustic,lai2011hybrid,fang2006ultrasonic,jing2015soft}. Dipole resonance originates by virtue of polarized oscillation with lower order rotational symmetry. Besides, monopole and quadrupole resonances 
emanate with non polarized oscillation with higher order rotational symmetry. 
Therefore, altering the various resonances to render the desired elastic wave bandgap through spatial manipulation of phases within the metamaterial is the new field of interest. 

The exploration of originally proposed  locally resonant metamaterial began with multiphase constituents for the evolution of negative effective mass at the dipole resonance~\citep{liu2000locally}. 
Later, harnessing multipolar resonances with the incorporation of multiple phases, additional bandgaps were obtained; however, bandgaps appear dispersed and small in width~\citep{li2004double,lai2011hybrid,wu2011elastic,sang2018design}.
Subsequently, the range of inclusive bandgap frequencies has been further increased by merging individual ones originating from dipole and monopole/quadrupole resonances~\citep{huang2012anomalous,lee2016origin,li2019modelling}. Nevertheless, the formation of bandgaps in metamaterial through this strategy essentially requires numerous materials leading to difficulty in fabricating multiple units with the appropriate configuration.
In order to overcome that problem, single phase metamaterials with multiple bandgaps have been evolved in recent years~\citep{zhu2014negative,warmuth2017single}.
In this regard, several  single phase cellular type structures have been proposed~\citep{jiang2021three,an20203d}, and particularly single phase star shape structure has been recognized for exhibiting an extended bandgap~\citep{chen2018band,chen2018design,meng2015band,li193multipolar}.
In parallel, recent experimental investigation on a single phase cellular structure also confirmed the existence of 11 kHz bandgap within 2 to 13 kHz frequency range~\citep{kumar2019low}.
Consequently, expanding the bandgap with minimalistic fabrication methodologies will have greater technological importance towards their applications in the sonic range.



\begin{figure}[!htb]  
	\centering
	\includegraphics[width=0.45\textwidth]{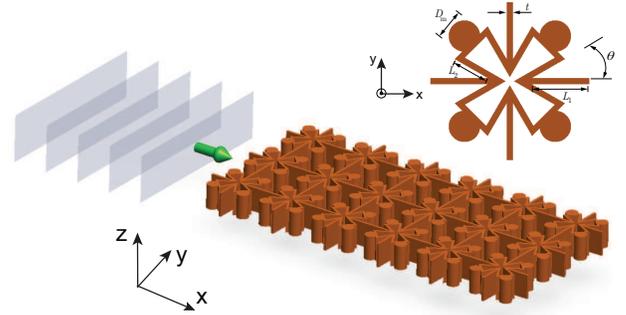}
	\caption{Schematic representation of periodic structure consisting of unit cells with in-plane wavefront. The Inset presents detail geometrical features.}
	\label{MetamaterialStructure}
\end{figure}

In this article, we systematically investigate the elastic wave bandgap of the metamaterial by localizing the same constituent material in the form of cylindrical mass at specific location as shown in Fig.~\ref{MetamaterialStructure}.
The unit cells are made of 3D printable resin (Visijet M3 Crystal) having material properties as Young's modulus~$E$=$1.463\,GPa$, Poisson's ratio~$\upsilon$=$0.37$, and density~$\rho$=$1020\,kg/m^3$.

\begin{figure}[]  
	\centering
	\subfloat{\includegraphics[width=0.26\textwidth]{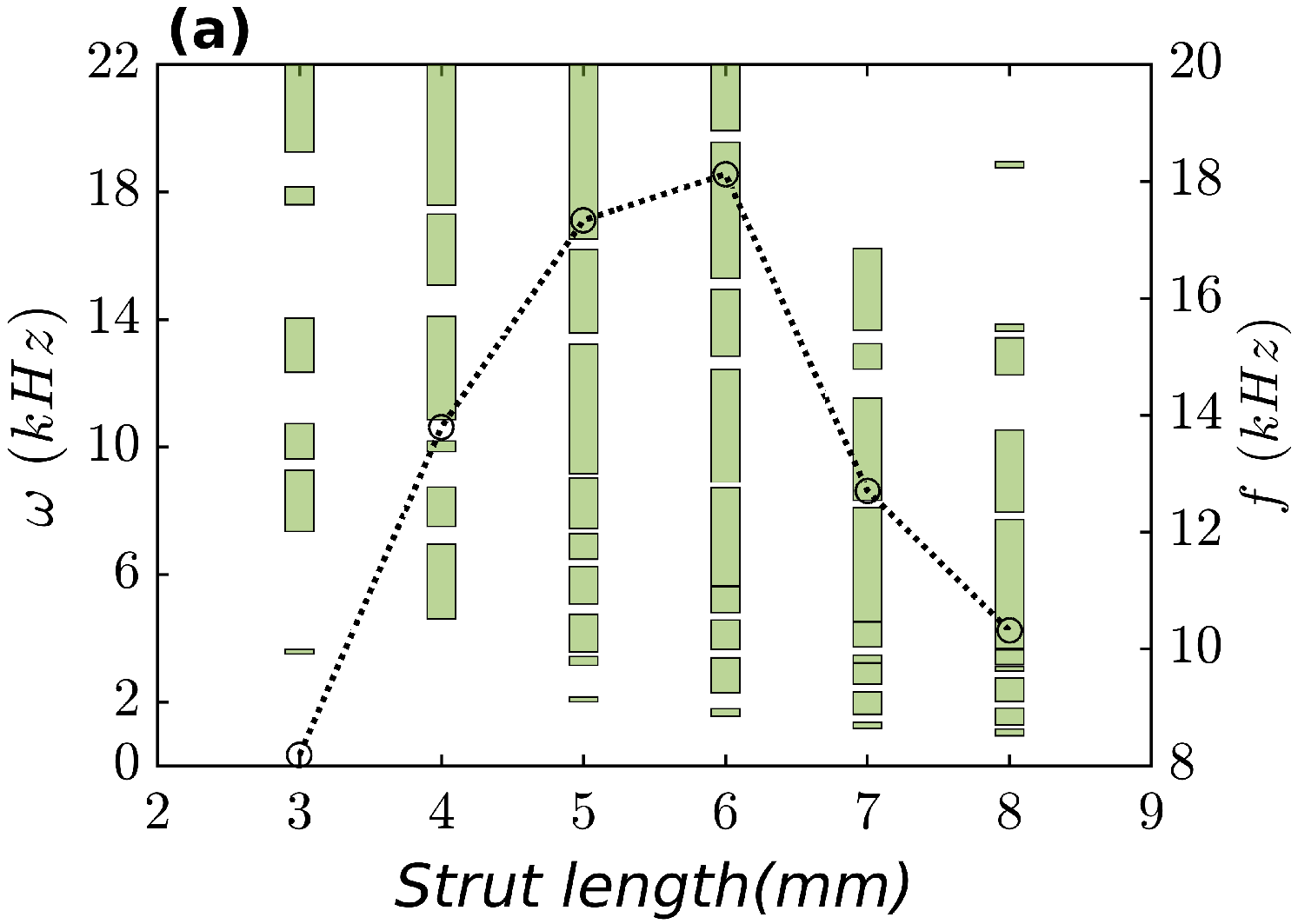}}
	\subfloat{\includegraphics[width=0.25\textwidth]{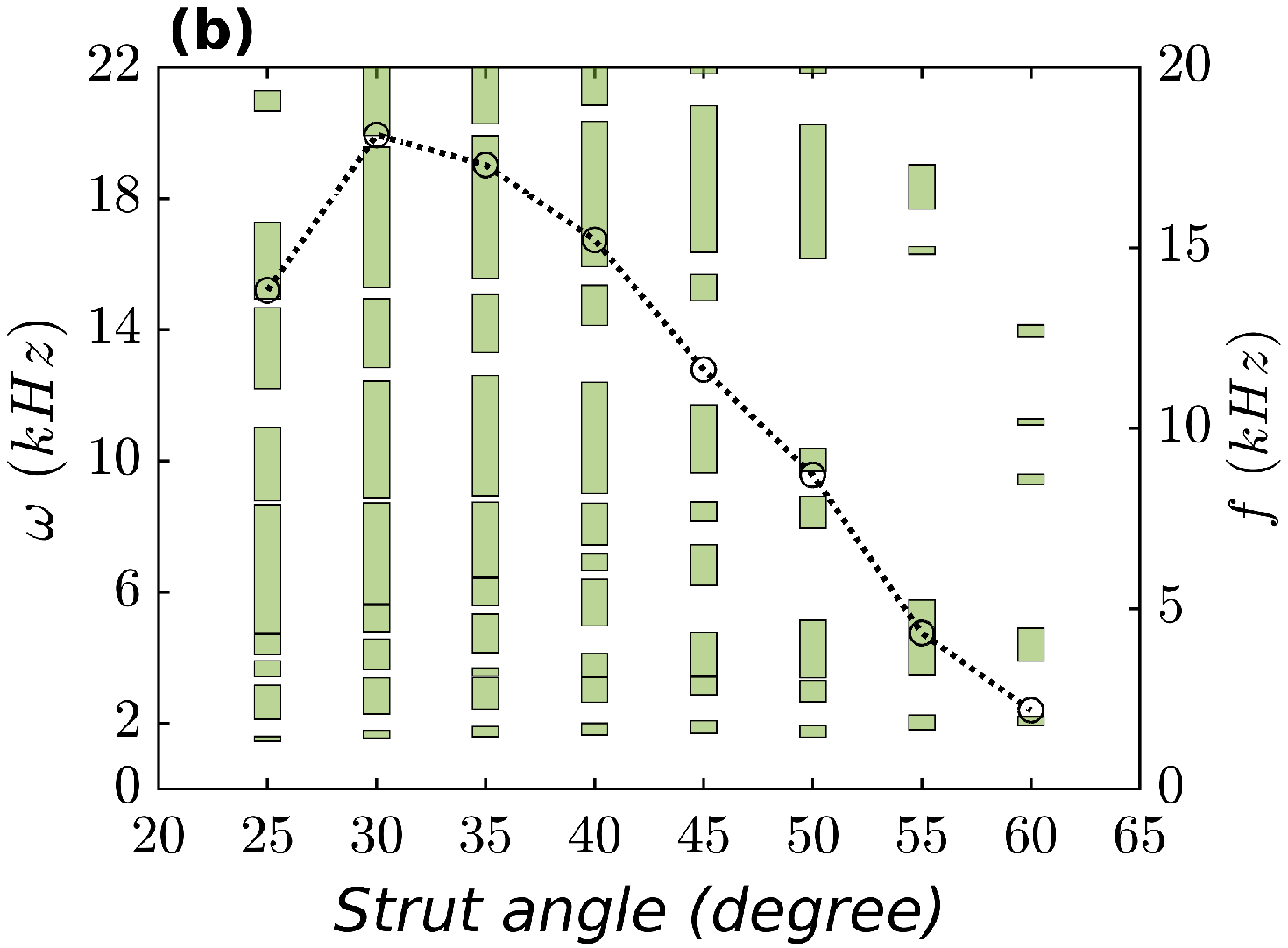}}
	\caption{Bandgaps and effective bandgap width vs struts length ($L_2$) and angle ($\theta$).}
	\label{fig:ParametricVariation}	
\end{figure}

The key geometrical features such as length of struts ($L_2$), angle between adjacent cell wall ($\theta$), and diameter of the cylindrical mass ($D_m$)  are considered to alter the emergent bandgaps.
Thereafter, the dispersion responses are simulated for the length of  struts with localized mass from $L_2$ = $3$ to $8\,mm$, while
keeping $L_1$, $\theta$, $D_{m}$, and $t$ constant
(see {\color{blue}Fig. S2} of {\color{blue}Supplementary Information (SI)}). Fig.~\ref{fig:ParametricVariation}(a) presents the bandgaps (green zones) variation and 
effective bandgap width (black dotted line) with the length of struts $L_2$. 
Effective bandgap width is calculated as  $f(\Delta \omega )$=$\sum_{i = 1}^{n} {(\omega _u^i}  - \omega _l^i)$, where $\omega_u^i$ and $\omega_l^i$ indicate upper and lower limits of $i^{th}$ bandgap among $n$ bands. 
It is inferred that interrupted bandgaps exist until $L_2$=4$\,mm$ and following addition of the length, bandgaps become more connected leading to an expanded frequency bandgap from 2 to 22 kHz. 
Upper limits exhibit 
almost 22 kHz till $L_2$=6$\,mm$; however, bandgap vanishes in high frequency range with sequential increments
of the length. Due to enlargement of the length  $L_2$, bending stiffness of the struts decreases, and overall mass increases. Because of this, the local resonance shifts in a lower frequency regime.
This fact eventually contributes to the appearance of a lower limit of the bandgap from 7.35 kHz to 1.27 kHz for $L_2$= 3 to $8\,mm$. It is observed that an eminent effective bandgap width greater than 17 kHz  is achieved only for $L_{2}$=5 to 6$\,mm$.
Hence, a metastructure  with $L_{2}$=6$\,mm$ reveals an overall bandgap of 18 kHz which 
is mostly undivided and as well as wider in low frequency regime.

Analogous to the previous section, we examine the variation of bandgaps and effective bandgap width versus the strut angle 
($\theta$)  
as indicated in Fig.~\ref{fig:ParametricVariation}(b) 
for the length, $L_2$=$6\,mm$. The dispersion responses for various strut angles are simulated and delineated in {\color {blue} Fig. S3} of {\color {blue} SI}.
It is recognized that 
uninterrupted and wide bandgaps arise till $\theta$=$40^{o}$. The subsequent change in angle to $\theta$=$45^{o}$,
several disconnected bandgaps occur. Apart from this, increment of angle above $\theta$=$45^{o}$ provides discontinuous bandgaps and overall bandgap diminish
from high and low frequency regimes. 
It is noticed that for the unit cell with localized mass united to the struts,
the effective bandgap width $f$ initially increases, and reaches an optimum value of 18.12 kHz for $\theta$=$30^o$.
After successive enhancement of strut angle, effective bandgap width $f$ linearly reduces and 
finally attains to 2.18 kHz at $\theta$=$60^{o}$. Similarly, we have varied the diameter of embedded mass associated with the struts for $\theta$=$30^o$ and $L_2$=$6\,mm$, and accordingly, detailed dispersion responses have been
presented in {\color {blue} Fig. S4} of {\color {blue} SI}. In correspondence to that, alteration of bandgaps and effective bandgap width against the diameter of embedded mass 
($D_{m}$)  have been depicted in  {\color {blue} Fig. S5} of {\color {blue} SI}.
Therefore, in 
this article, the unit cell with $\theta$=$30^o$, and $L_2$=$6\,mm$ along with localized mass $D_{m}$= $5\,mm$ attached to the struts delivers a low frequency and continuous wide bandgap.

\begin{figure}[!htb]  
	\centering
	\setlength\abovecaptionskip{-0.6\baselineskip}
	\subfloat{\includegraphics[width=0.5\textwidth]{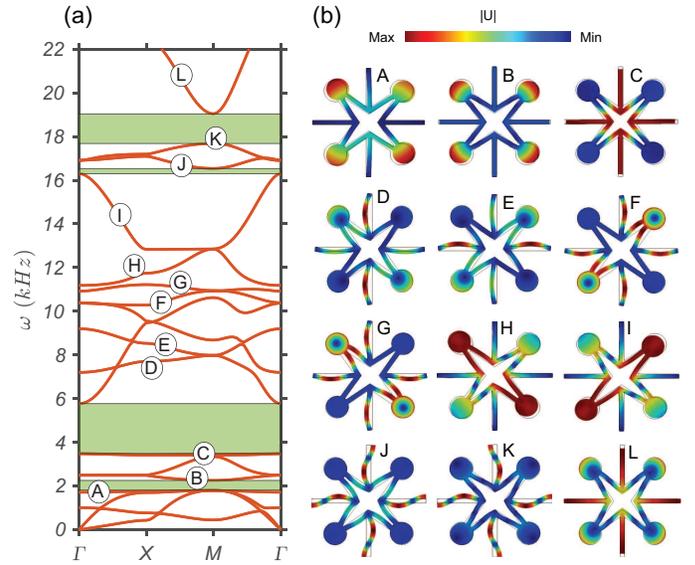}}
	\caption{(a) Dispersion responses of the metastructure with strut angle $\theta$=$55^o$, and (b) Bloch mode shapes for alphabetically marked bands.}
	\label{fig:Bandgapresultspecimen1}	
\end{figure}

\begin{figure}[!htb]  
	\centering
	\setlength\abovecaptionskip{-0.6\baselineskip}
	\subfloat{\includegraphics[width=0.5\textwidth]{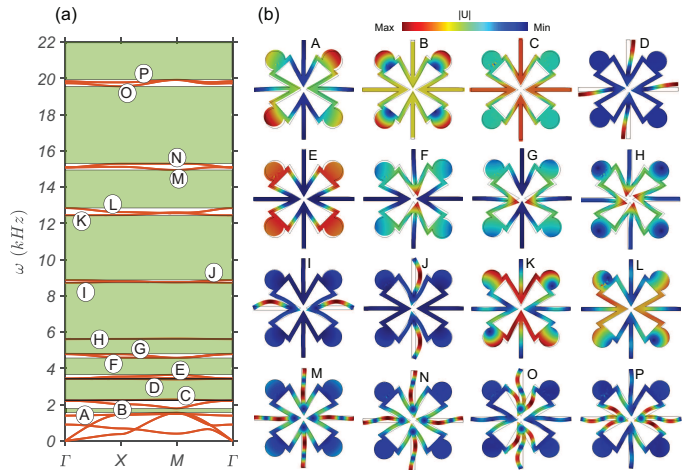}}
	\caption{(a) Dispersion responses of the metastructure with strut angle $\theta$=$30^o$, and (b) Bloch mode shapes for alphabetically marked bands.}
	\label{fig:Bandgapresultspecimen2}	
\end{figure}

Based on the presence of interrupted and consecutive bandgaps, we have chosen two metastructures with $\theta$=$55^o$ and $\theta$=$30^o$, respectively. Primarily, we will evaluate the bandgap numerically for these structures and compare them with the experimentally obtained mechanical wave transmission. In conjunction, the origin of this bandgap will be illustrated with associated Bloch modes for these two structures.
We first investigate the band structure of the unit cell with localized mass for $\theta$=$55^o$ in which discontinuous bandgaps develop in 1.8 to 2.25 kHz, 3.5 to 5.8 kHz, and 17.7 to 19 kHz as shown in Fig.~\ref{fig:Bandgapresultspecimen1}(a). Interestingly, in this structure directional bandgaps ($\Gamma$-$X$ direction) are evident  from 1.8 to 5.8 kHz, 10.4 to 10.9 kHz, 11.75 to 12.8 kHz, 16.3 to 16.9 kHz, and 17.2 to 22 kHz (see {\color {blue}SI} in Fig. S8(a)). To understand the band structure, as well as the formation of the bandgaps, Bloch modes of bands A to L are extracted at $M$ point and displayed in Fig.~\ref{fig:Bandgapresultspecimen1}(b). In order to explore the formation of directional bandgaps, Bloch modes are calculated at $X$ point (see {\color {blue}SI} in Fig. S8(b)).
The deformed Bloch modes are overlaid to the undeformed geometry outlined with black color. Band A explains the rotation of diagonal struts in reverse directions arising in bending of vertical struts and no deformation in horizontal struts reveals the dipole resonances~\citep{kumar2019low,liu2000locally}.  Besides, band B implies compression and extension of horizontal and vertical struts consisting of opposite rotation of struts with embedded mass. 
In particular, this vibration mode clearly marks the evolution of quadrupole resonance due to the existence of lower order (two-fold) rotational symmetry. Explicitly, quadrupole resonance is analogous to a higher order dipole composed of two alike dipoles with opposite phases~\citep{russell1999acoustic,kumar2020tunable}. Later, band C demonstrates the inward movement of horizontal and vertical struts accommodating contraction of struts with localized mass. This mode shows the higher order (four-fold) rotational symmetry which signifies the evolution of monopole resonance~\citep{kumar2019low,lai2011hybrid}. 
Afterwards, band D displays the bending of horizontal and vertical struts which cause the deformation in two diagonal struts while other diagonal struts remain stationary.
Notably, the vibration modes of band D and band E depict similar deformation with different orientations. Therefore, these modes contribute the same effective properties (negative effective mass and modulus) ensuing in no bandgap at $M$ point~\citep{wang2004two,sun2010resonant,wang2014harnessing,kumar2021unraveling}.
Likewise, the bandgap is absent at $M$ point between the vibration modes of bands F and G as well as bands H and I. 
Noticeably for band J, outward and inward deformation of struts in a diagonal direction and bending of vertical and horizontal struts deliver the quadrupole resonance.
In succession, band K exhibiting rotation of struts with localized mass in an anticlockwise direction, and bending of horizontal and vertical struts portrays the monopole resonance.
Meanwhile, band L indicating compression and expansion of horizontal and vertical ribs along with bending deformation of diagonal struts delineates quadrupole resonance. Conclusively, fewer bands
acquire the dipole resonance in the lower frequency regime leading to negative effective mass while several bands witness monopole and quadrupole resonances in higher frequency regimes resulting in negative effective modulus.  

In addition to the prior analysis, the band structure of the unit cell with cylindrical mass attached to the struts with angle $\theta$=$30^o$ in which mostly consecutive bandgaps appear from 2 to 22 kHz is shown in Fig.~\ref{fig:Bandgapresultspecimen2}(a). Dispersion response comprises of bandgap from 1.56 to 1.79 kHz, 2.29 to 3.4 kHz, 3.65 to 4.57 kHz, 4.8 to 5.6 kHz, 5.6 to 8.72 kHz, 8.88 to 12.44 kHz, 12.85 to 14.95 kHz, 15.3 to 19.5 kHz, and 19.9 to 22 kHz. As compared to the previous case, the bandgaps are extended for both high and low frequency regimes.
Specifically, the 
accumulative bandgap of 18 kHz originates with considering all small pass-bands. To find out the bandgap formation, Bloch modes of bands A to P are extracted at $M$ point and expressed in Fig.~\ref{fig:Bandgapresultspecimen2}(b). Band A demonstrates the rotation of struts with embedded mass in the reverse direction arising in bending of horizontal struts, and no deformation in vertical struts preserves the dipole resonances. Thereafter, the band B imparts extension and compression of horizontal and vertical struts having an opposite rotation of diagonal struts. Thus, it clearly declares the presence of quadrupole resonances. Further, band C presents the outward movement of horizontal and vertical struts accommodating the expansion of diagonal struts and reveals monopole resonance. In contrast to the previous case $\theta$=$55^o$, the magnitude of deformation is higher for bands B and C.
Thus, the locally resonant unit amplifies the exertion of net force to the structure, and accordingly, these bands interact with the high frequency bands~\citep{wang2004two,wang2014harnessing,kumar2021unraveling}.
In subsequent, the band D states the bending of four straight struts in the opposite direction while core (struts tied to the mass) prevails undeformed which offers quadrupole resonance. Analogous to band D, band E denotes the rotation of struts with localized mass in the reverse direction while straight struts persist stationary. Besides, band F and band G show alike deformation with different phases, such that no bandgap is generated at the $M$ point. Later, the band H prescribes the twisting of the core in a clockwise direction while no deformation is marked in straight struts clearly indicating the torsional mode. In this mode, deformation is limited to the core and hence no force is exerted to the structure by the core. Consequently, due to the lack of its interaction with the external wave field, no bandgap is evolved at the $M$ point~\citep{wang2004two,krushynska2014towards}.
Afterwards, bands I and J manifest the deformation of two straight struts while
other struts remain straight appearing in dispersion response corresponding to the
standing waves as noticed in conventional LRMs. 
Thus, vibration modes of bands I and J are interpreted as the emergence of stationary cores while the soft coating is subjected to
rotation under dynamic conditions~\citep{krushynska2014towards,kumar2019low}. 
Furthermore, bands K and L suggest almost identical deformation with distinct phases indicating no substantial bandgap between these bands at the $M$ point.
Nonetheless, band M offers a mode shape with four fold rotational symmetry pointing to monopole resonance. Meanwhile, band N delineates the inward and outward deformation of diagonal struts with localized mass and bending of four straight struts which convey the signature of quadrupole resonance. 
The band O is occasioned by significant bending of vertical struts which causes the deformation in one side of diagonal struts and dictates the quadrupole resonance. 
Likewise, it is evident that band P exhibits a similar quadrupole of band O however in a different phase. Thus, no bandgap is formed at the $M$ point between bands O and P. In conclusion, the development of strong monopole and quadrupole resonances and their subsequent cooperation with high frequency bands for the metastructure with $\theta$=$30^o$ leads to the formation of wide and continuous bandgaps.

\begin{figure}[]  
	\centering
	\subfloat{\includegraphics[width=0.5\textwidth]{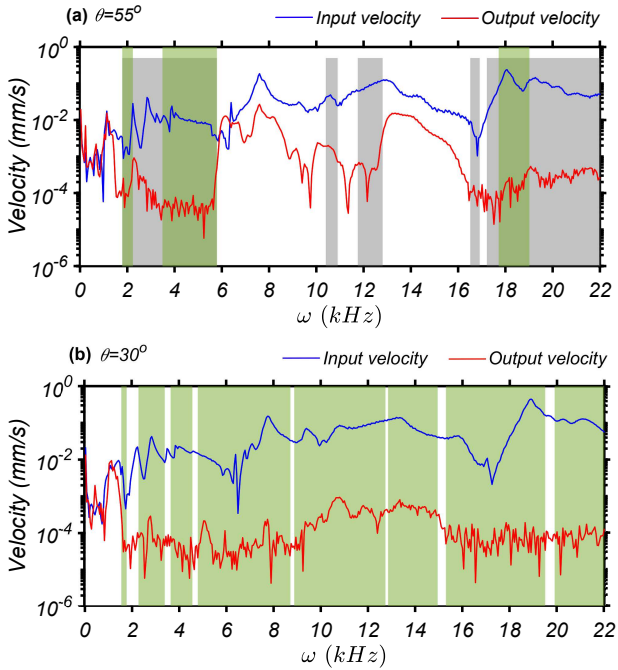}}
	\caption{Experimentally measured input and output velocity signals vs frequency for the metastructure with (a) $\theta$=$55^o$, and (b) $\theta$=$30^o$. Green and gray color zones indicate the omnidirectional ($\Gamma$-$X$-$M$-$\Gamma$ path) and directional ($\Gamma$-$X$ path) bandgaps, respectively.}
	\label{fig:ExperimentalResult}	
\end{figure}

We further investigate the experimental transmission spectra with few unit cells of fabricated metastructure for $\theta$=$55^o$ and $\theta$=$30^o$ (for detailed experimental procedure see the {\color {blue}SI}). In the present study, using a function generator, a sine wave is produced through a piezo actuator by sweeping the frequency from 50 Hz to 22 kHz over 0.02 seconds with an interval of 50 Hz. 
Utilizing, Laser Doppler Vibrometer, the incident mechanical signal (velocity signal) is extracted on the aluminum bar and the output signal is collected on the left (free) end of the specimen. Both the input and output velocity signals are contrasted with numerically determined bandgaps as shown in Fig.~\ref{fig:ExperimentalResult}.
The comparison of the incident and transmitted velocity signals for the metastructure with $\theta$=$55^o$ is described in Fig.~\ref{fig:ExperimentalResult}(a), where green and gray zones denote the numerically predicted omnidirectional ($\Gamma$-$X$-$M$-$\Gamma$ path) and directional ($\Gamma$-$X$ path) bandgaps, respectively. We examine the transmission of the elastic wave over numerically evaluated directional bandgaps from 1.8 to 5.8 kHz, 10.4 to 10.9 kHz, 11.75 to 12.8 kHz, 16.3 to 16.9 kHz, and 17.2 to 22 kHz frequency ranges.
It is observed that the
bandgaps marked by the gray zone lie within the frequency regime where the output signal is remarkably attenuated as compared to the input signal. 
In parallel, Fig.~\ref{fig:ExperimentalResult}(b) delineates the incident and transmitted velocity signals for the metastructure with $\theta$=$30^o$ along with its omnidirectional bandgaps represented by green zones. Although the metastructure $\theta$=$30^o$ shows considerable transmission in the 0 to 1.8 kHz range, it drastically weakens ($\sim100$ times) in the elevated frequency range (1.8 to 22 kHz).
The reduction of the signal amplitude over this range is persistent with the numerical prediction. Thus, the 
experimentally inspected attenuation in transmission for $\theta$=$30^o$ is in good agreement to the omnidirectional bandgaps, while the attenuation for $\theta$=$55^o$ is in accordance with numerically obtained directional bandgaps.  

\begin{figure}[]  
	\centering
	\subfloat{\includegraphics[width=0.5\textwidth]{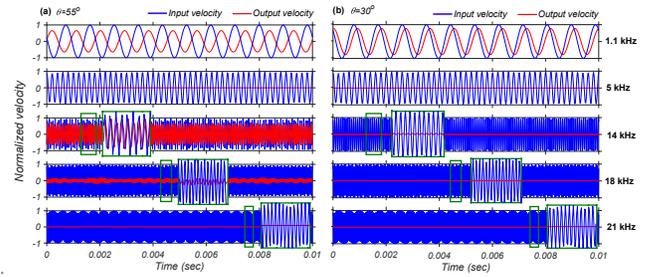}}
	\caption{Comparison of experimentally obtained input and output velocity signals 
		for the metastructure at different frequencies with struts angle (a) $\theta$=$55^o$, and (b) $\theta$=$30^o$. Insets compare the zoomed incident and transmitted signals.}
	\label{fig:Bandgapillustration}	
\end{figure}


We further carried out the experiment to analyze the transmission with sine signals of fixed frequency. Of particular interest is to substantiate the presence of stop and pass bands. Accordingly, we investigate the incident and transmitted velocity signals for both the metastructures ($\theta$=$55^o$, and  $\theta$=$30^o$) as illustrated in Fig.~\ref{fig:Bandgapillustration}(a) and (b), respectively.  In particular, to precisely examine the transmitted signals, we normalize them with the peak value of the incident velocity signal. Based on the experimental investigation (Fig.~\ref{fig:ExperimentalResult}), we choose the individual sine signals at 1.1 kHz, 5 kHz, 14 kHz, 18 kHz, and 21 kHz for both of the configurations. We identify the significant transmission with respect to the incident signal at 1.1 kHz for both the metastructures as appeared in the first row of Fig.~\ref{fig:Bandgapillustration}(a) and (b). 
Thus, the frequency of 1.1 kHz lies in the pass band as noted in Fig.~\ref{fig:ExperimentalResult}(a) and (b).
Thereafter, the second row of Fig.~\ref{fig:Bandgapillustration}(a) and (b)
demonstrate almost no transmission for the metastructures at 5 kHz, and the observed results are consistent with the stop bands as found in Fig.~\ref{fig:ExperimentalResult}(a) and (b). Notably, a velocity signal with 14 kHz shows the pronounced transmission for metastructure of angle $\theta$=$55^o$. In contrast, metastructure with angle $\theta$=$30^o$ denotes no transmission at 14 kHz. Thus, the output velocity signals of both the metastructures at 14 kHz are persistent with the recognized transmissions in Fig.~\ref{fig:ExperimentalResult}. 
Moreover, as expected, transmitted velocity signals of both the metastructures at frequency 18 kHz represent the negligible transmission as stated in the fourth row of Fig.~\ref{fig:Bandgapillustration}(a) and (b).
Likewise, transmitted values at 21 kHz (fifth row of Fig.~\ref{fig:Bandgapillustration}(a) and (b)) prove no propagation of waves within the metastructures. This is due to the directional bandgap for $\theta$=$55^o$ as displayed in Fig.~\ref{fig:ExperimentalResult}(a). 
Altogether, the detailed inspections of signal transmissions at specific frequencies further support the existence of omnidirectional bandgaps in $\theta$=$30^o$ as well as directional bandgaps as exhibited by $\theta$=$55^o$.


In summary, the localized mass embedded metastructure with angle $\theta$=$30^o$ delivers an overall bandgap of 18 kHz in the 2 to 22 kHz frequency range, while $\theta$=$55^o$ provides only 5 kHz in the same range. Further, experimentally obtained attenuation in mechanical wave transmission spectra for metastructures over the same frequency range closely match with the numerically predicted bandgaps.~Additionally, a comparison of the incident and transmitted velocity signals for specifically chosen frequencies further supports the distinct bandgaps for the metastructures. Detailed analyses of Bloch wave modes certify the presence of strong monopole and quadrupole resonances towards wide and connected bandgaps.~Notably, the metastructure with angle $\theta$=$30^o$ yields omnidirectional bandgaps, while the metastructure with $\theta$=$55^o$ evolves several directional bandgaps. Finally, the present investigation reveals a minimalistic reconfiguration strategy for achieving the enlarged frequency bandgap in single phase cellular material.
\\
\\
{\color {blue}The supplementary material} consists of metamaterial geometry,  
a numerical framework to predict the dispersion response, formation of bandgaps and its associated Bloch modes for different geometrical parameters, and an
experimental procedure  to obtain the transmission spectra for the fabricated metastructures.


The authors greatly acknowledge all support from SERB-DST Grant Number 
ECR/2016/001635 and are also thankful to Tinkering Lab for fabrication facilities.

%

\renewcommand{\thefigure}{S\arabic{figure}}
\renewcommand{\theequation}{S\arabic{equation}}
\setcounter{figure}{0}
\section*{Supporting Information}
This supporting information contains the evolution of elastic wave bandgap for various geometrical parameters in single phase lattice type metamaterials. Employing the finite element procedure, dispersion responses are extracted within the irreducible Brillouin zone (IBZ). Further, selecting the key design parameters such as length of struts, angle between adjacent cell walls, and diameter of localized mass, bandgaps are illustrated. Moreover,
underlying mechanisms towards the formation of bandgap are demonstrated by identifying the local resonances such as dipole, monopole, and quadrupole resonances. Finally, the experimental transmission spectra for elastic wave propagation within the metastructure are obtained and compared with numerical predictions.
\section{Geometry of the metamaterial and numerical framework}
\begin{figure}[!htb]  
	\centering
	\includegraphics[width=0.5\textwidth]{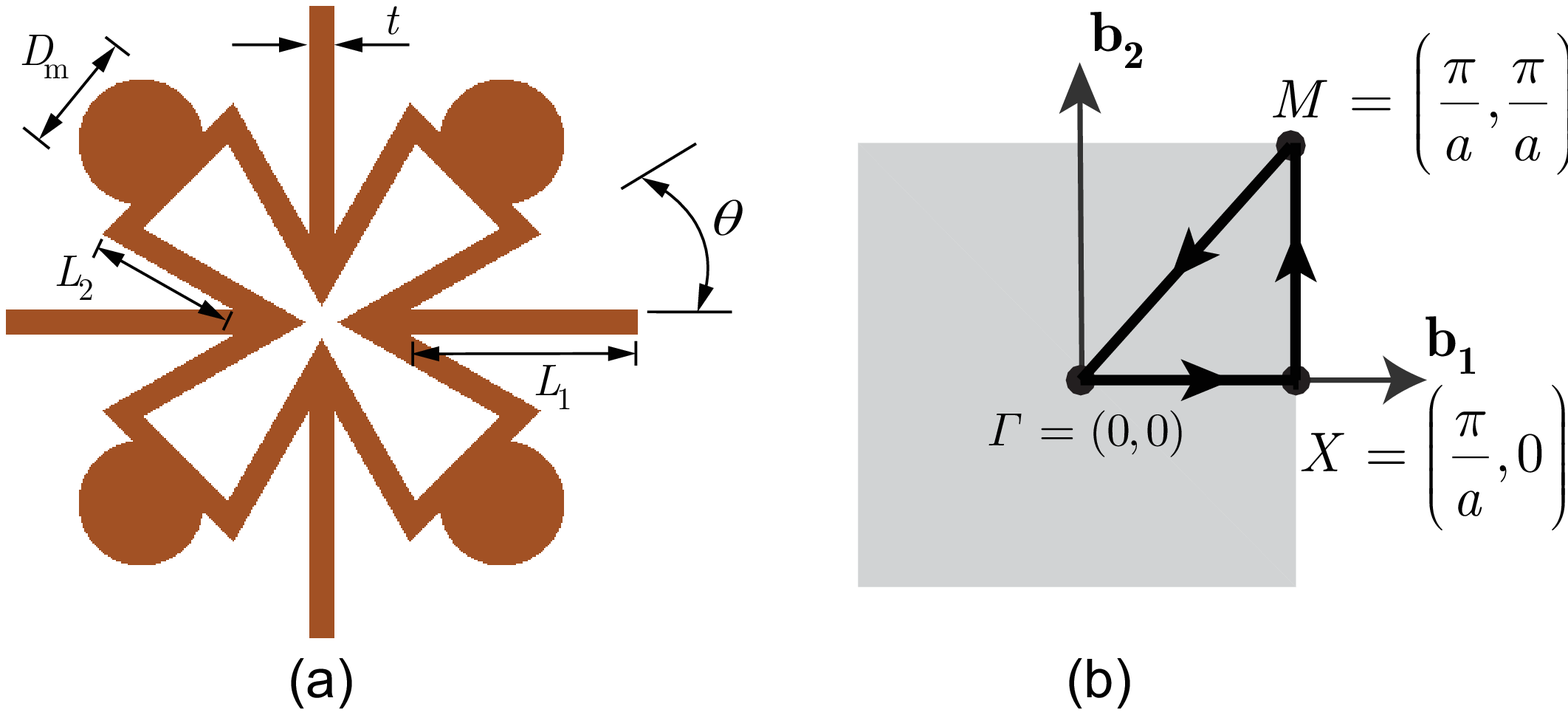}
	\caption{Unit cells with localized mass connected to the struts with angle $\theta$=$30^o$, and (b) Irreducible Brillouin Zone~(IBZ) in reciprocal lattice bounded by $\Gamma$-$X$-$M$-$\Gamma$.
	}
	\label{fig:Geometries}
\end{figure}

In Fig.~\ref{fig:Geometries}(a), we describe the geometrical parameters of the unit cell for the metastructure as presented in Fig. 1 of the main article.
It is composed of eight struts of equal length $L_2$=$6\,mm$ with diameter of cylinder $D_{m}$=$5\,mm$, angle between adjacent cell wall $\theta$=$30^o$, four straight struts $L_1$=$9\,mm$, and thickness of the structure  $t$=$1\,mm$. We have taken the unit cell size $a$=$25\,mm$. The wave vector $\mathbf k=(k_{x}, k_{y})$
along the boundary of the IBZ (Fig.~\ref{fig:Geometries}(b)) will be used to find the dispersion curve.
According to the classical Bloch-Floquet theorem~\citep{kittel1976introduction}, 
the displacement field ${\mathbf u}$ at location ${\mathbf r_{o}}$
can be presented as, 
\begin{equation}
{\mathbf u (\mathbf r_{o},t)}={\tilde{\mathbf u}}_{{\mathbf k} }(\mathbf r_{o}) e^{i({\mathbf k}\cdot{\mathbf r_{o}}-\omega t)} 
\label{eq:dispfield}
\end{equation}
where, ${\mathbf k}$ denotes the wave vector, $i$=$\sqrt{-1}$ represents the imaginary number, and $\omega$ is the angular frequency. 
Here, magnitude of the lattice periodic displacement field is denoted by ${\tilde {\mathbf u}}_{{\mathbf k}}(\mathbf r_{o})$, which is invariant under lattice translation  {\bf{T}}. Afterwards, using Bloch-Floquet theorem, the displacement field is expressed as 
\begin{equation}
{\mathbf u ({\mathbf r_{o}}+{\mathbf T},t)}={\mathbf u ({\mathbf r_{o}},t)}e^{i {\mathbf k }\cdot {\mathbf T}}
\label{eq:bloch}
\end{equation}

The above equation provides the kinematic constraint which ensures ${\mathbf u }$ be periodic. 
Elastodynamic wave propagation in a deformable continuum
without body force can be noted as
\begin{equation}
\nabla  \cdot \boldsymbol{\sigma} = \rho \ddot{\mathbf u} 
\label{eq:equilibrium}
\end{equation}
where $\rho$ is the density of the material. Incorporating plane wave expansion of time dependent displacement field Eq.~(\ref{eq:dispfield}), Eq.~(\ref{eq:equilibrium}) can be formed as 
\begin{equation}
\nabla \cdot \boldsymbol{\sigma}  +  \rho {\omega ^2}{\mathbf u} = {\mathbf 0}
\label{eq:dynamic}
\end{equation}

Employing finite element procedure~\citep{kumar2019low} of the Eq.~(\ref{eq:dynamic}) along with the kinematic constraint (Eq.~(\ref{eq:bloch})), the eigen value problem in the reduced displacement field can be obtained as,
\begin{equation}
[{\mathbf K(\bf{k})}-{\omega^2}{\mathbf M(\bf{k})}]{\mathbf u}_r={\mathbf 0}
\label{eq:Eigen_equation}
\end{equation}
where,  $ {\bf{K}}$ and  
$ {\bf{M}}$
are stiffness and mass matrices, respectively. We use the Scalable Library for Eigenvalue Problem Computations (SLEPc)~\citep{campos2012slepc} package to solve the large scale eigen value problem. Moreover, the dispersion curve can be evaluated in the reciprocal lattice for any point ${\mathbf k}$=~$(k_{x},k_{y})$ by tracing the Bloch wave number in the IBZ as shown in  Fig.~\ref{fig:Geometries}(c).
At first, dispersion responses are calculated along the $\Gamma$-$X$-$M$-$\Gamma$ path by altering the length of struts $L_{2}$ from $3$ to $8\,mm$ as depicted in Fig.~\ref{fig:Bandgapresultlength}.
In an analogous manner, 
the bandgaps for various strut angles ($\theta$) and diameter of the localized mass ($D_{m}$) are 
also presented in Fig.~\ref{fig:Bandgapresultangle} and Fig.~\ref{fig:Bandgapresultdiameter}, respectively.

\section{Parametric investigation of bandgap}
\begin{figure*}[!htb]  
	\centering
	\subfloat[{}]{\includegraphics[width=4cm,height=8cm]{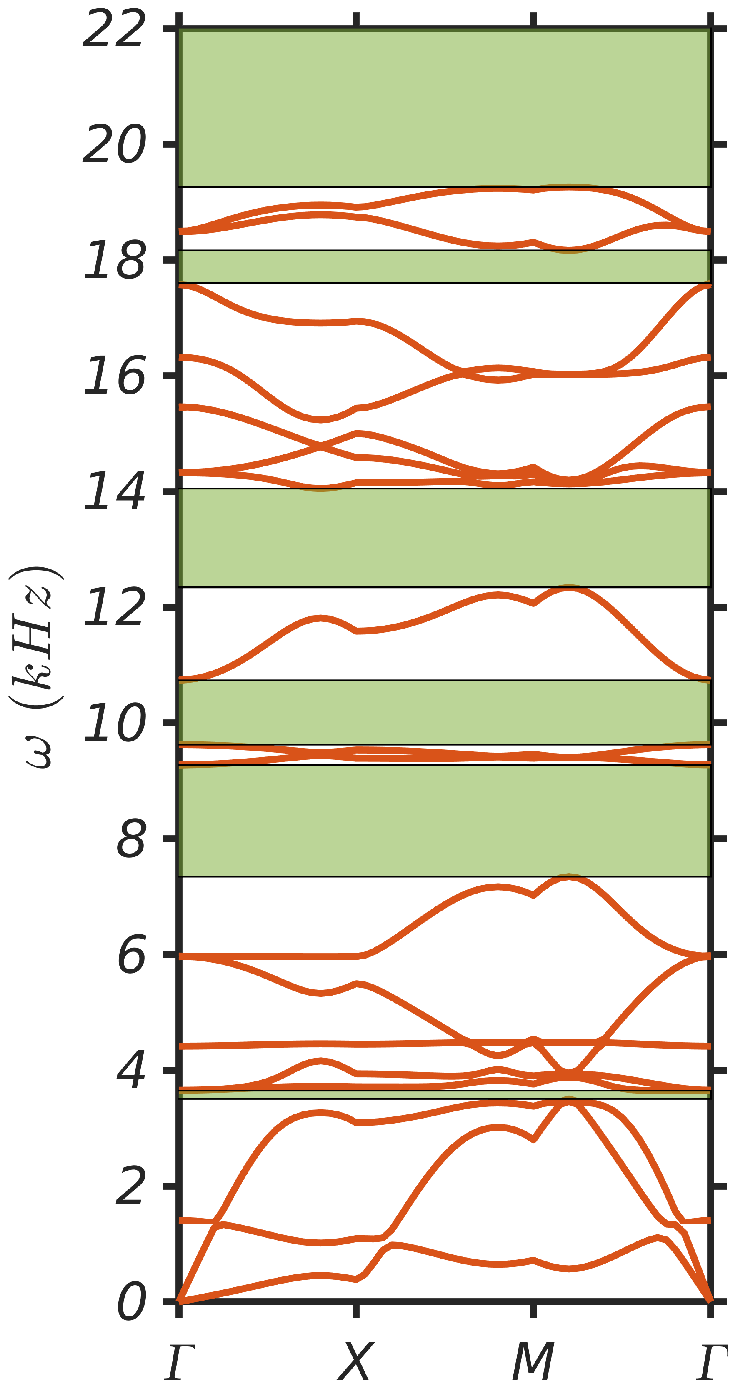}}
	\subfloat[{}]{\includegraphics[width=4cm,height=8cm]{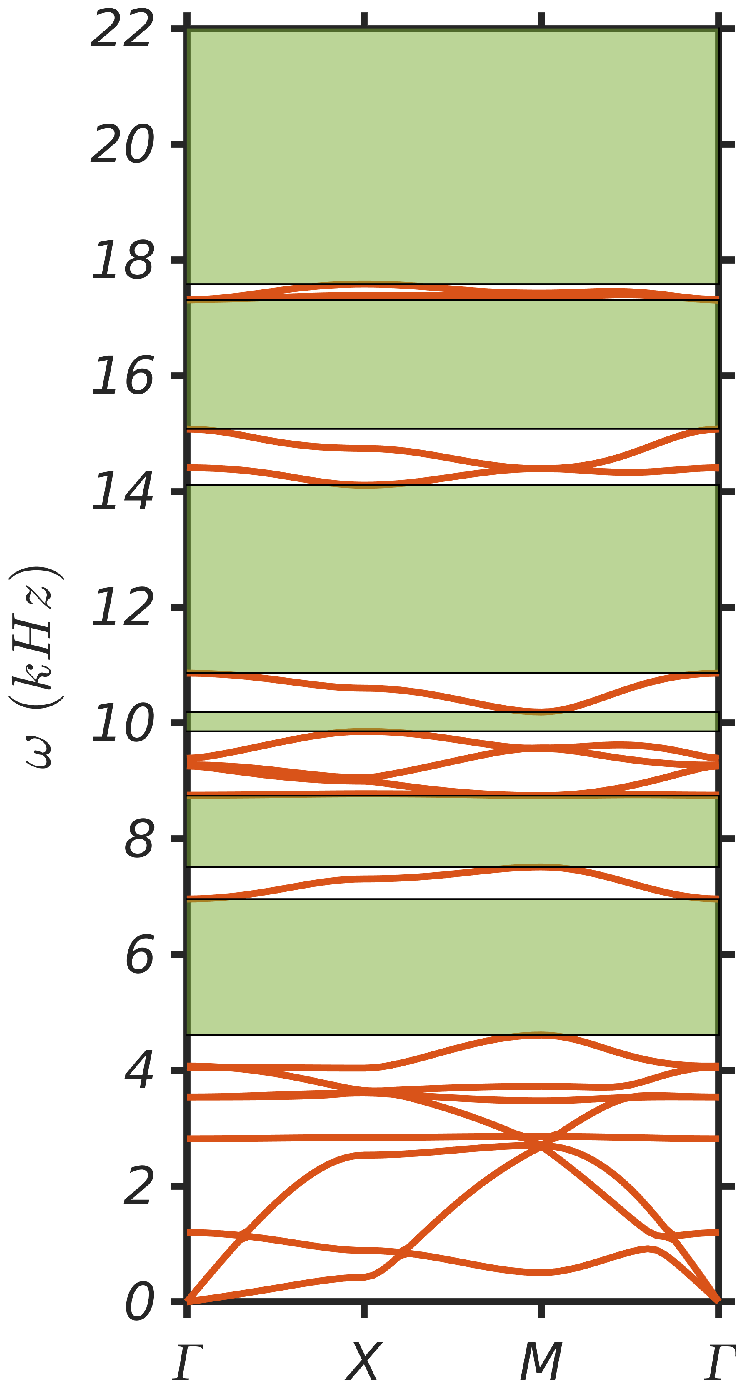}}
	\subfloat[{}]{\includegraphics[width=4cm,height=8cm]{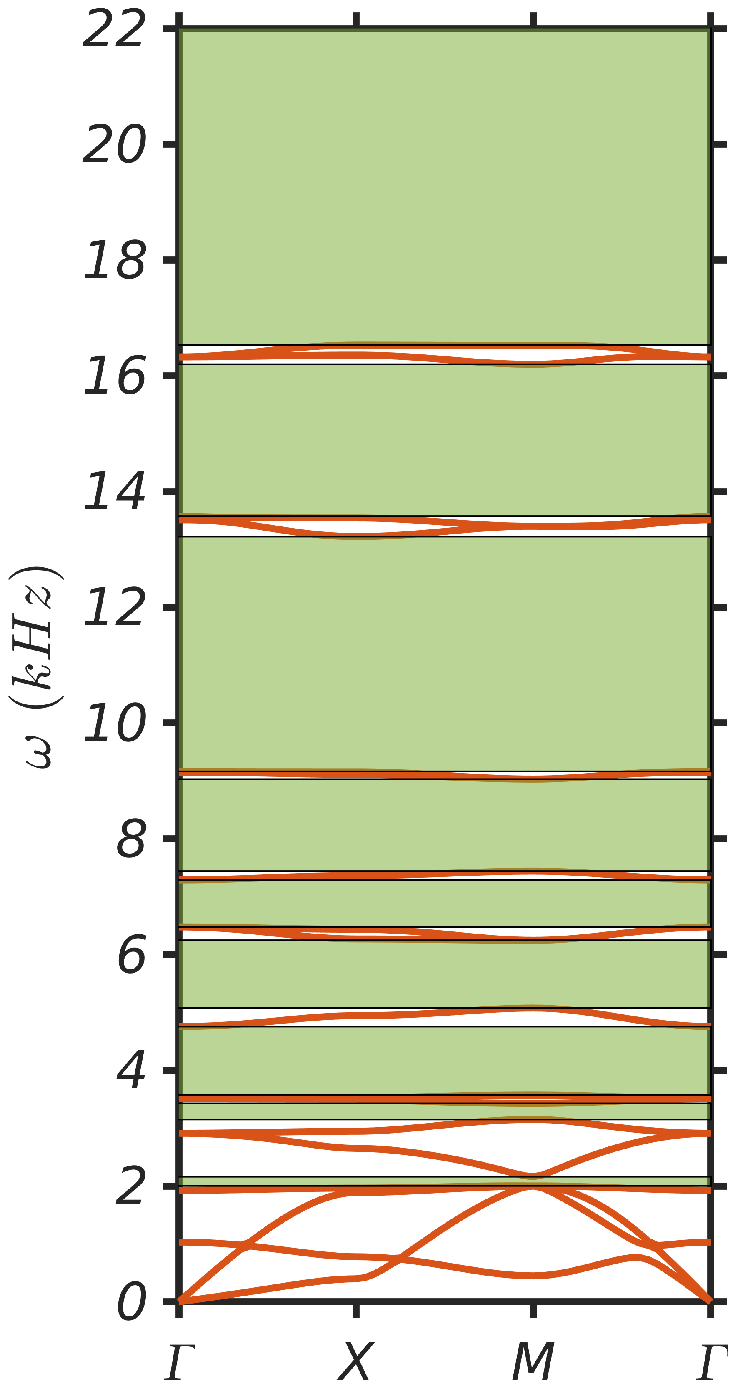}}\\
	\subfloat[{}]{\includegraphics[width=4cm,height=8cm]{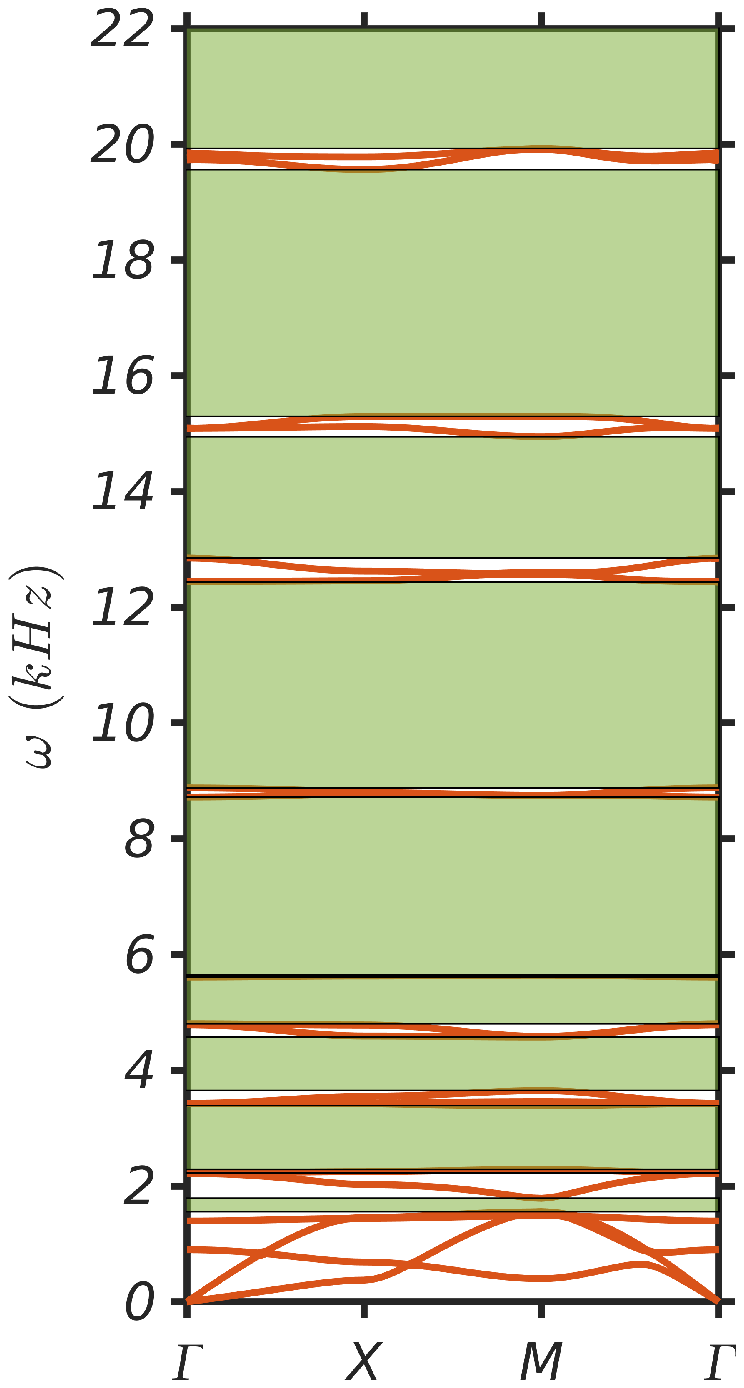}}
	\subfloat[{}]{\includegraphics[width=4cm,height=8cm]{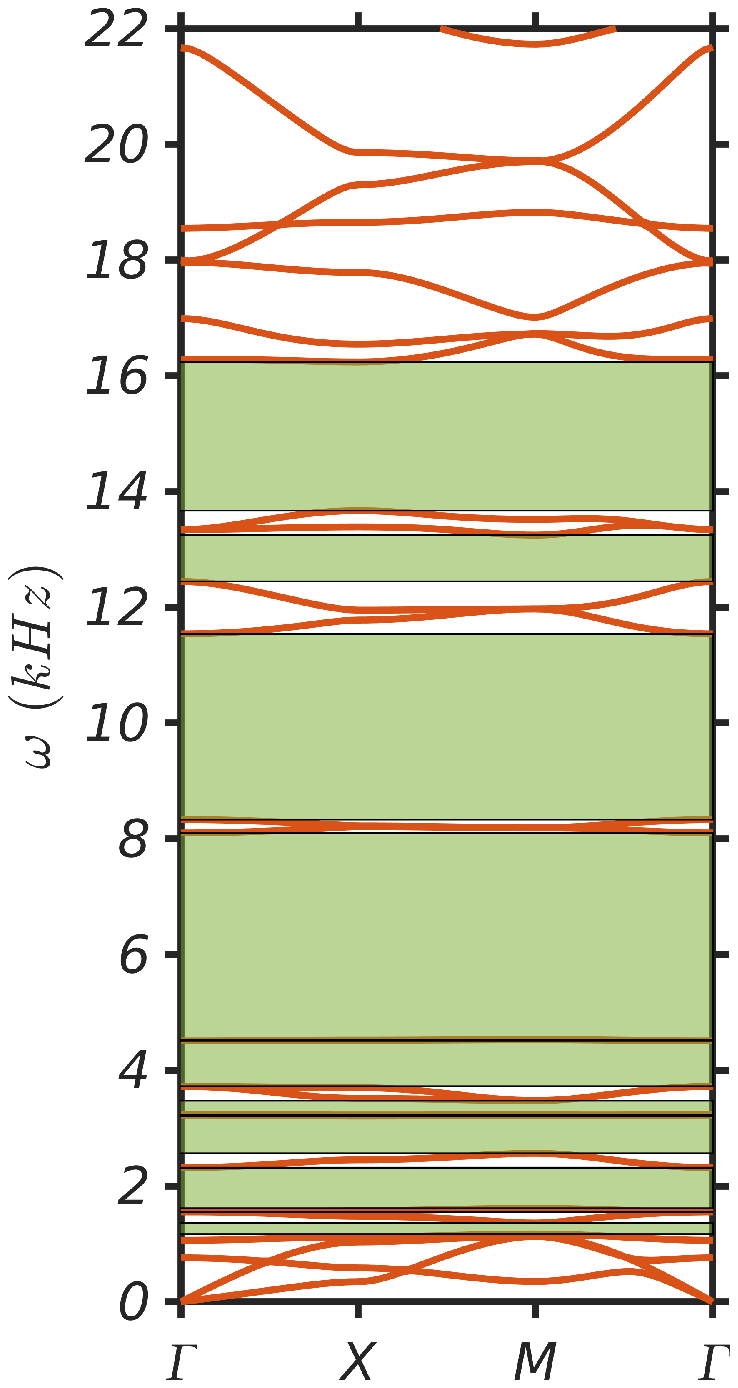}}
	\subfloat[{}]{\includegraphics[width=4cm,height=8cm]{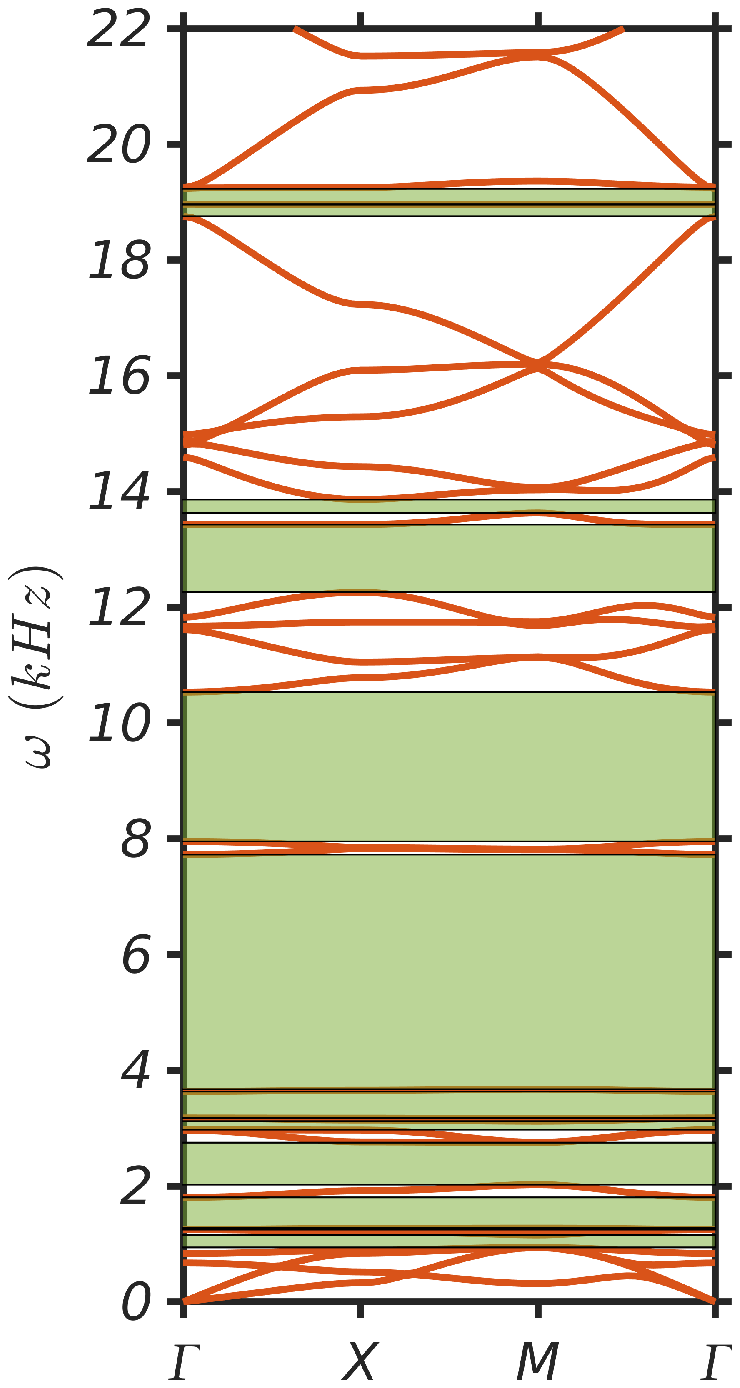}}
	\caption{Dispersion curves for struts length (a) $L_2$=$3\,mm$, (b) $L_2$=$4\,mm$, (c)  $L_2$=$5\,mm$,  (d)  $L_2$=$6\,mm$, (e)  $L_2$=$7\,mm$, and (f)  $L_2$=$8\,mm$ along $\Gamma$-$X$-$M$-$\Gamma$ path.}
	\label{fig:Bandgapresultlength}	
\end{figure*}

We perform a detailed parametric investigation to find the bandgaps by varying geometrical parameters for the metastructure.
The key design parameters which tailor the bandgaps are determined as the length of struts ($L_2$), struts angle ($\theta$), and diameter of localized mass ($D_{m}$). Alteration of bandgaps with different strut lengths is illustrated in  Fig.~\ref{fig:Bandgapresultlength}.
It is observed that when the length increases, the bending stiffness of struts decreases, and overall mass increases which enforce the bandgap to develop at a lower frequency regime. This fact demonstrates the shifting of local resonances in lower frequency regimes with a higher value of $L_2$.

\begin{figure*}[!htb]  
	\centering
	\subfloat[{}]{\includegraphics[width=4cm,height=8cm]{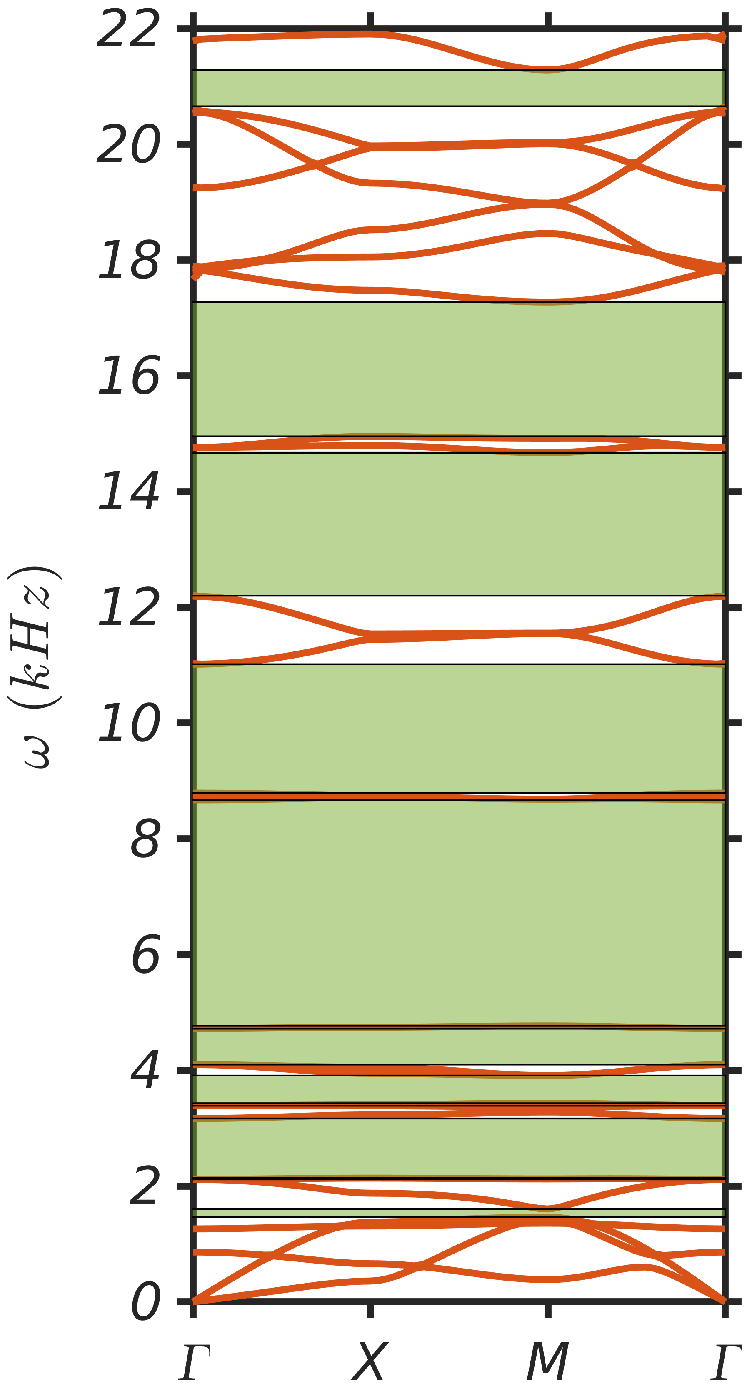}}
	\subfloat[{}]{\includegraphics[width=4cm,height=8cm]{STARN5L6.eps}}
	\subfloat[{}]{\includegraphics[width=4cm,height=8cm]{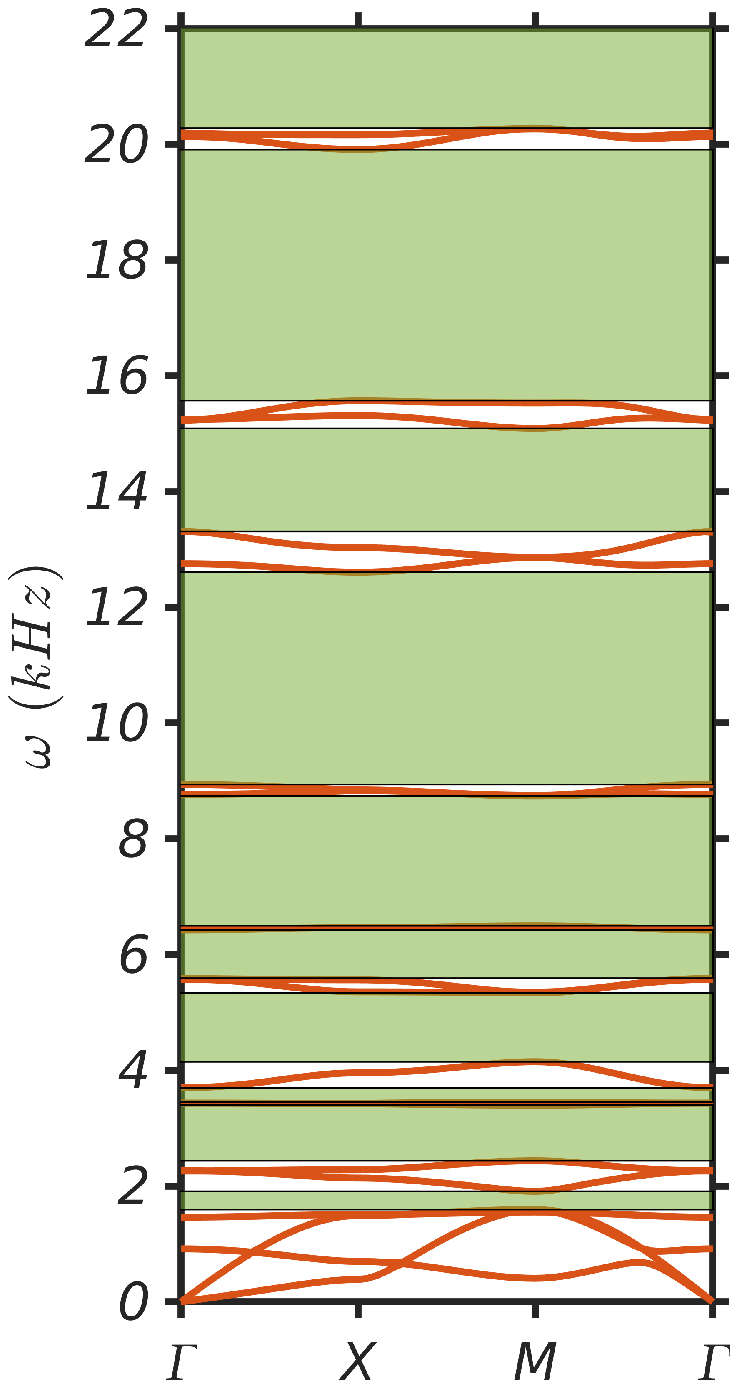}}
	\subfloat[{}]{\includegraphics[width=4cm,height=8cm]{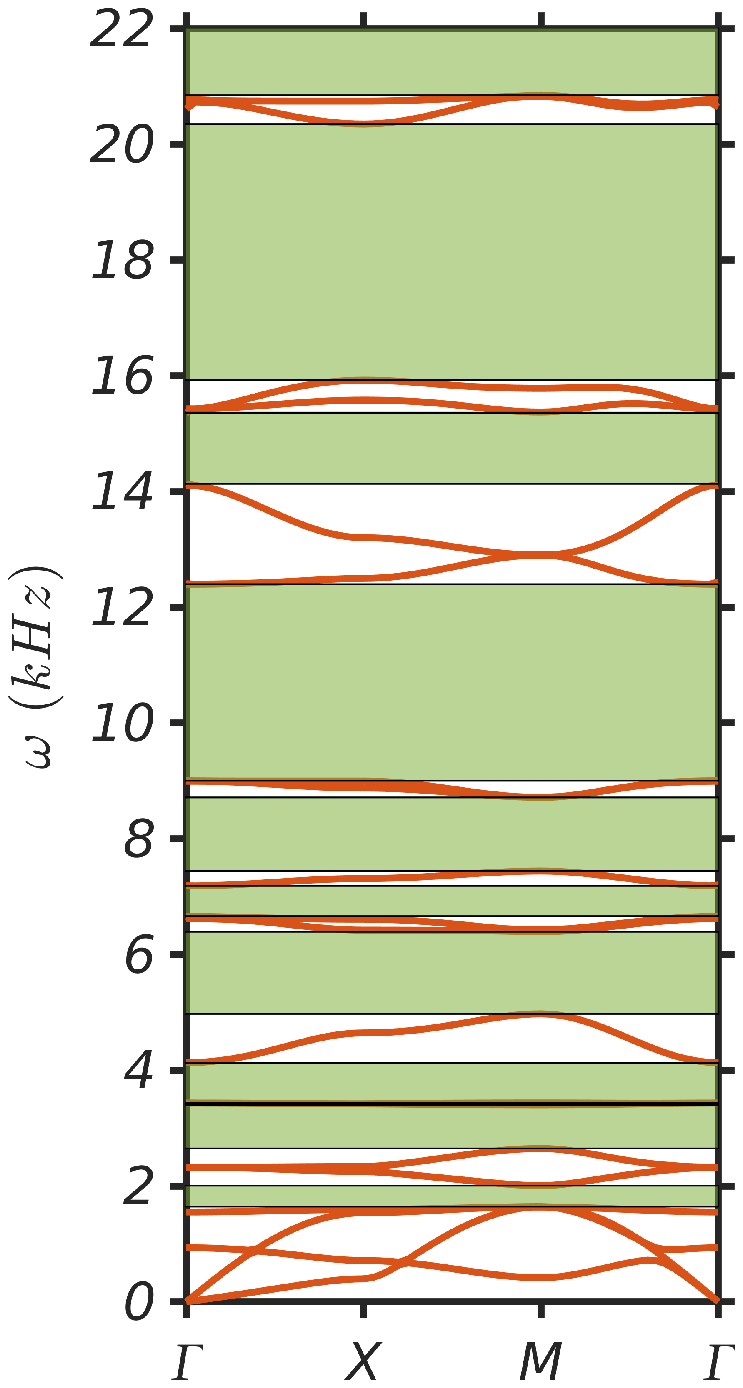}}\\
	\subfloat[{}]{\includegraphics[width=4cm,height=8cm]{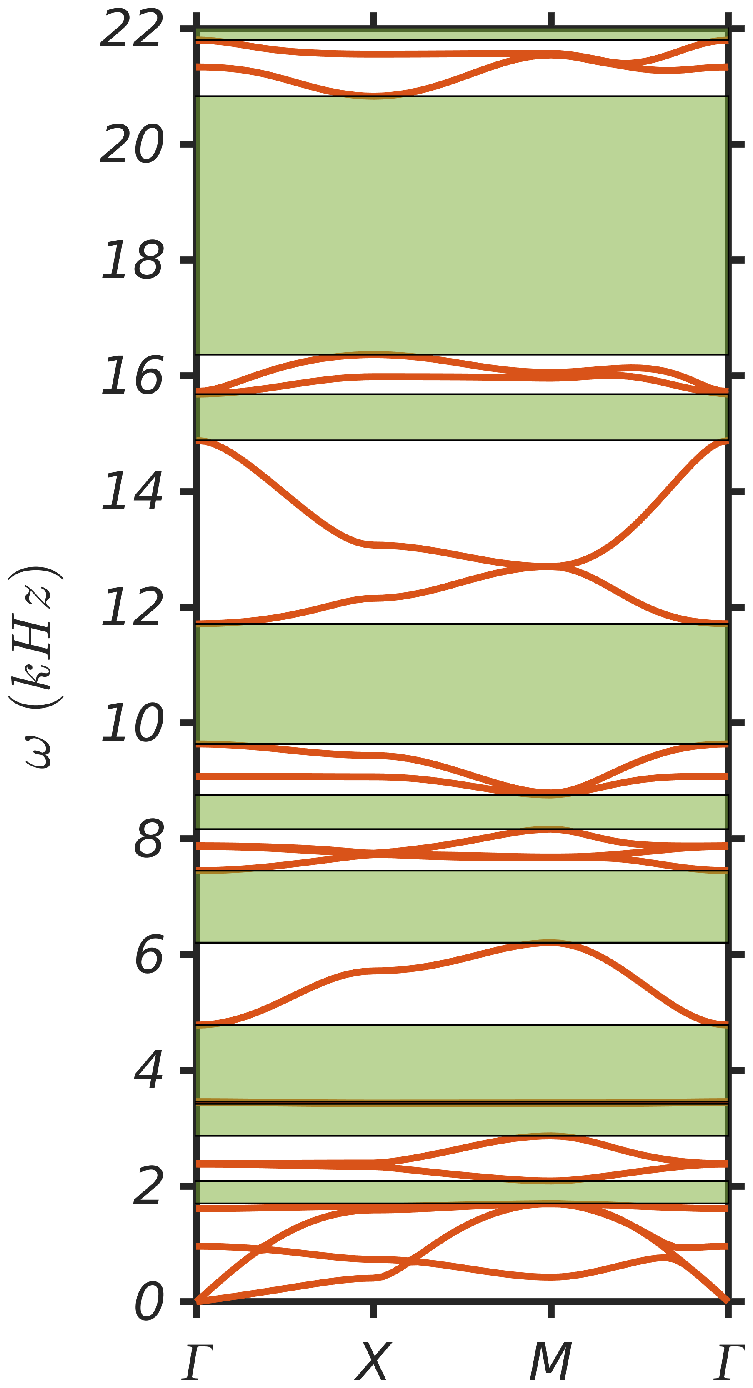}}
	\subfloat[{}]{\includegraphics[width=4cm,height=8cm]{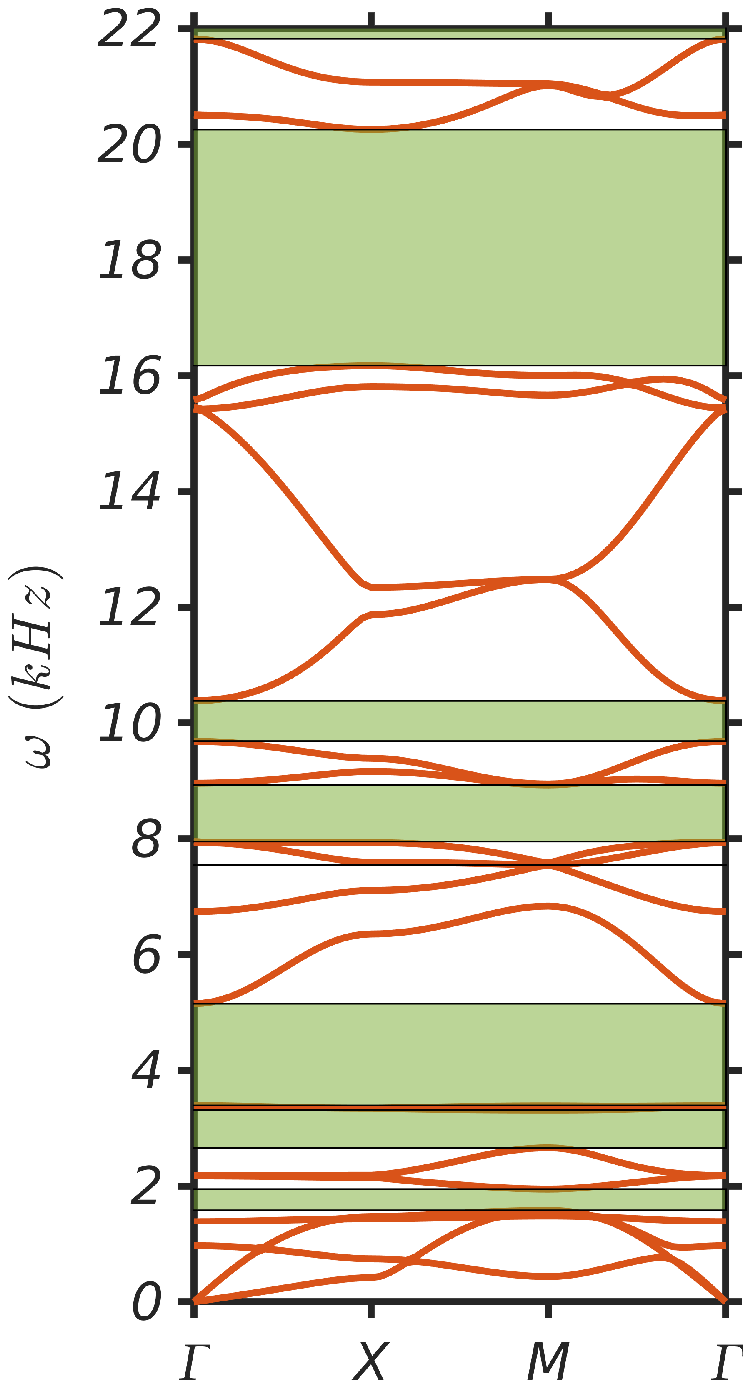}}
	\subfloat[{}]{\includegraphics[width=4cm,height=8cm]{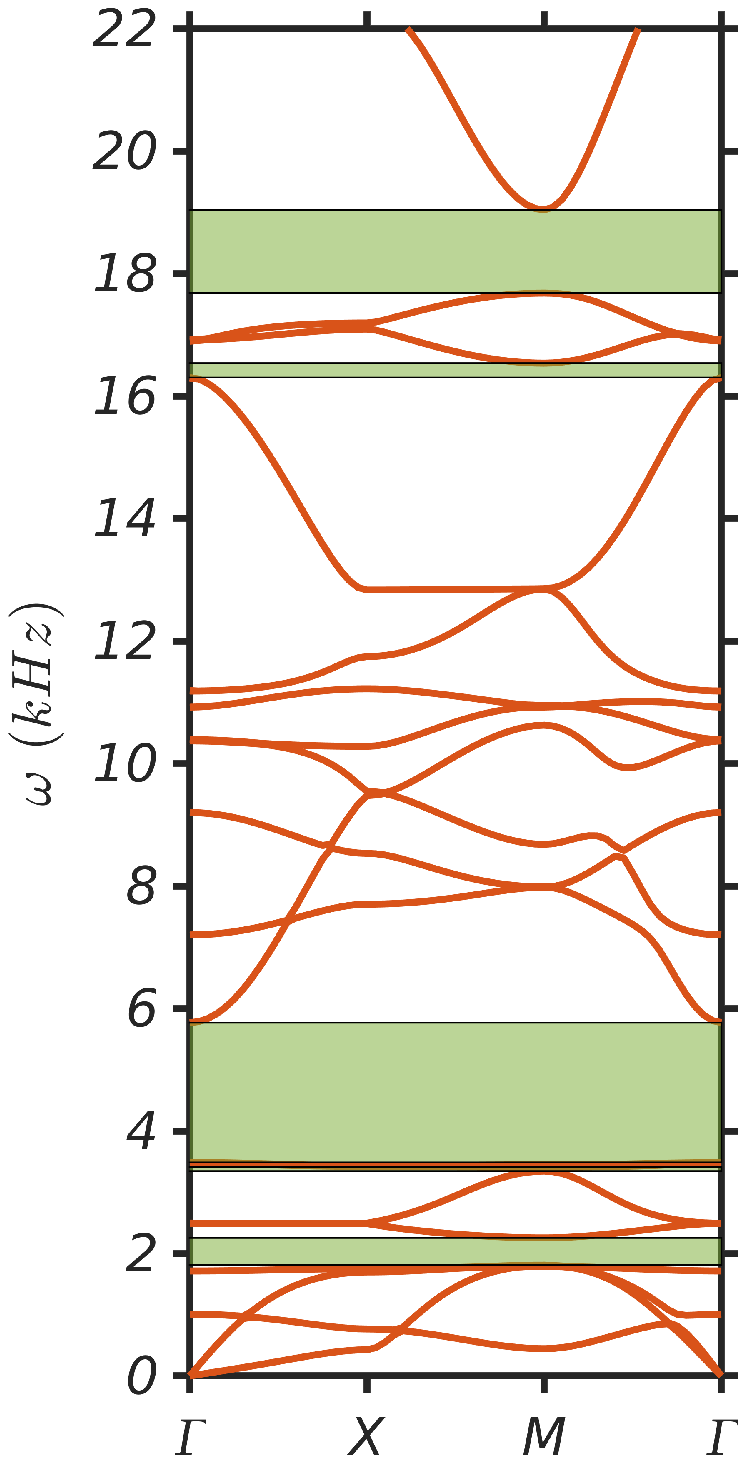}}
	\subfloat[{}]{\includegraphics[width=4cm,height=8cm]{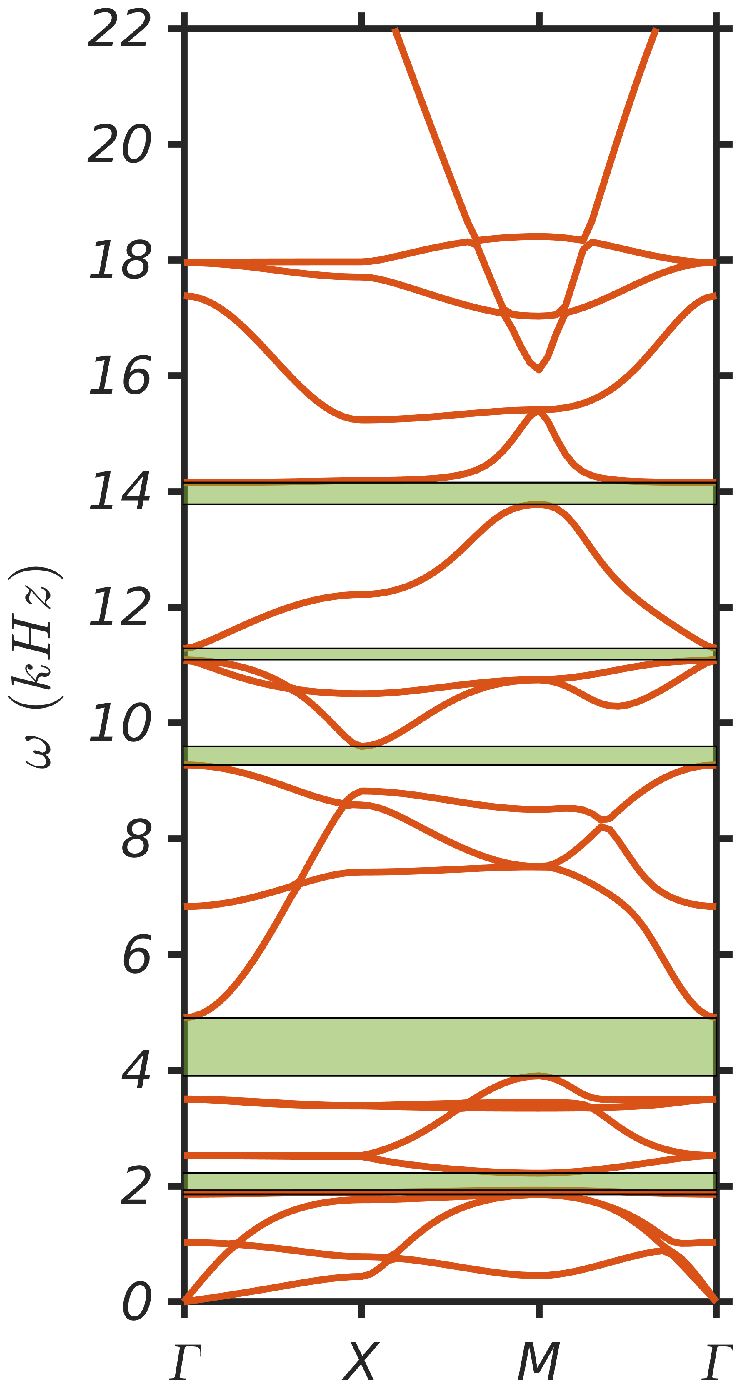}}
	\caption{Dispersion curves for different strut angle (a) $\theta$=$25^o$, (b) $\theta$=$30^o$, (c) $\theta$=$35^o$, (d) $\theta$=$40^o$, (e) $\theta$=$45^o$, (f) $\theta$=$50^o$, (g) $\theta$=$55^o$, and (h) $\theta$=$60^o$ along $\Gamma$-$X$-$M$-$\Gamma$ path.}
	\label{fig:Bandgapresultangle}	
\end{figure*}

In contrast, the struts angle plays an important role to change the negative effective properties. It is inferred from Fig.~\ref{fig:Bandgapresultangle} that with an increment of strut angle low frequency bandgap ($<$2 kHz) is elevated to the slightly higher frequency regime upon reduction of the static mass. Consequently, bandgaps above the 2 kHz frequency regime significantly diminish with the decrease of strut angle. 

\begin{figure*}[!htb]  
	\centering
	\subfloat[{}]{\includegraphics[width=4cm,height=8cm]{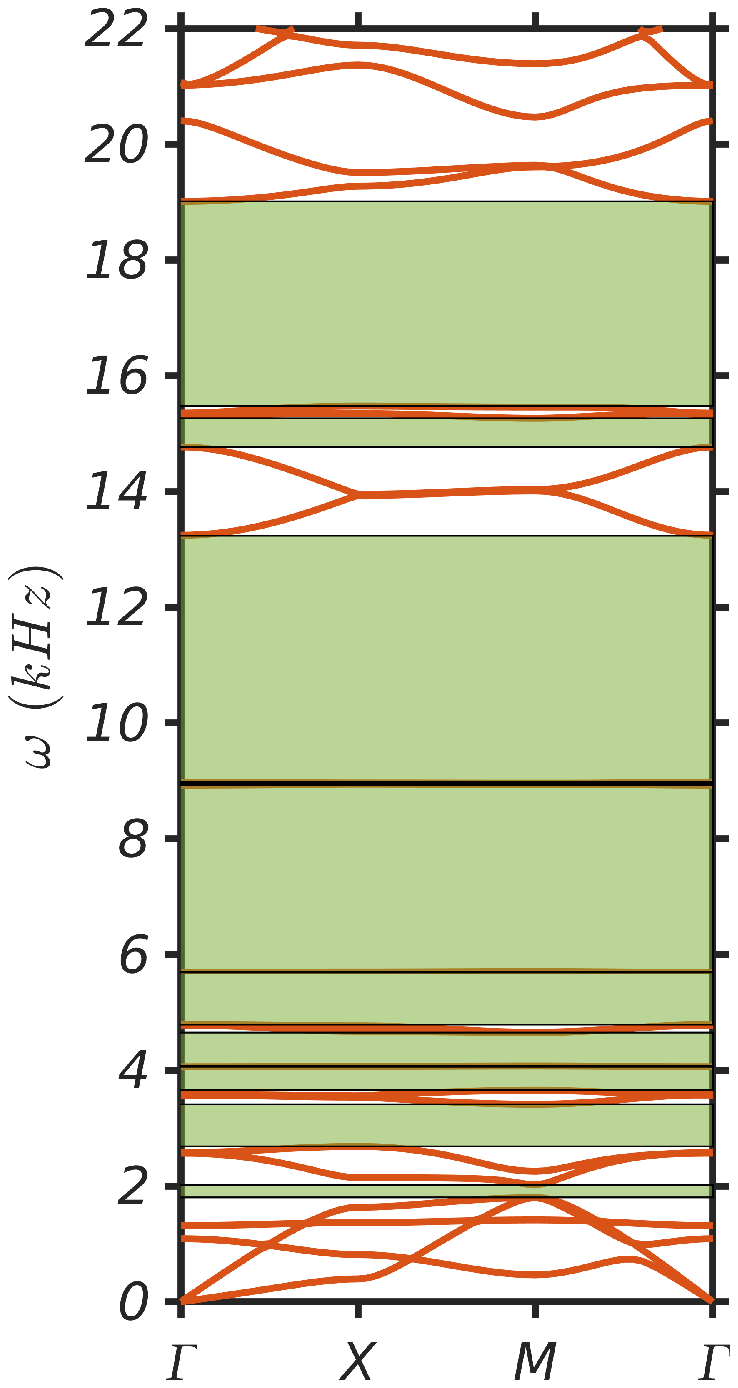}}
	\subfloat[{}]{\includegraphics[width=4cm,height=8cm]{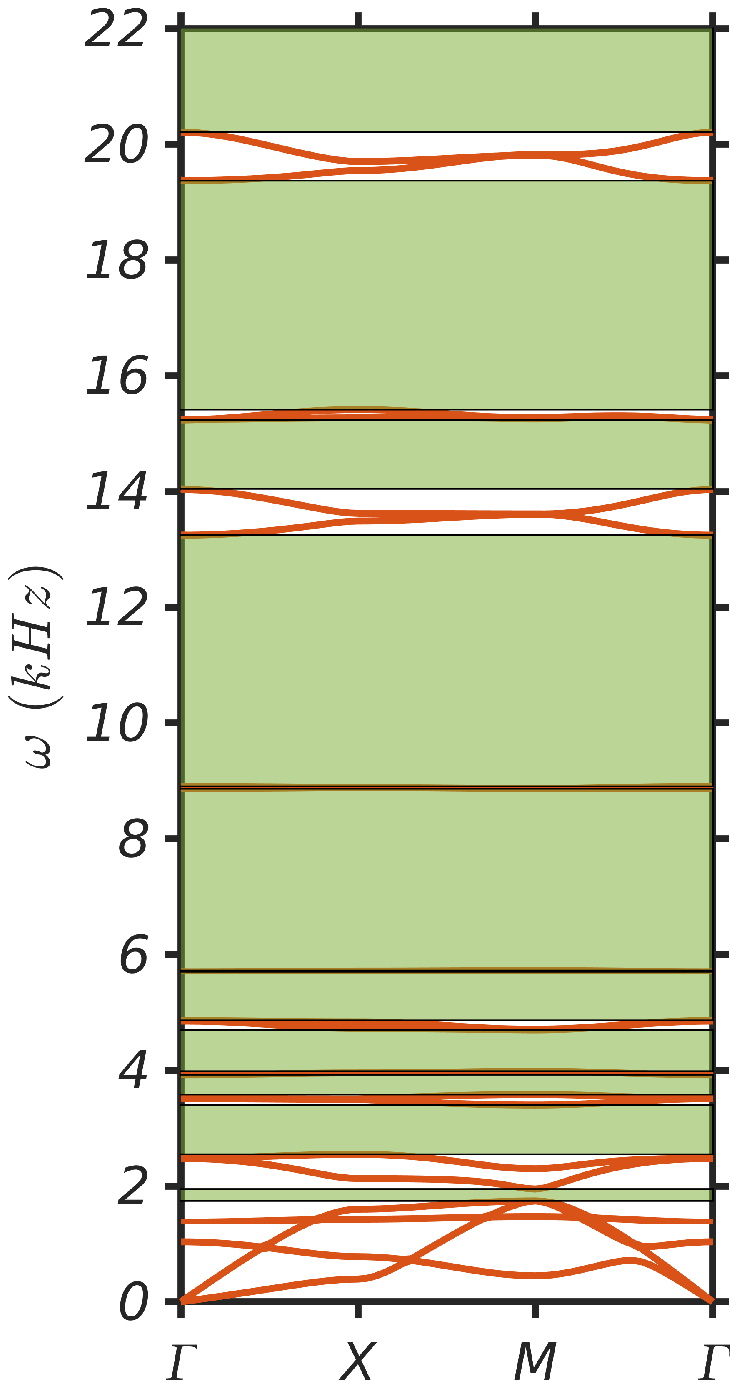}}
	\subfloat[{}]{\includegraphics[width=4cm,height=8cm]{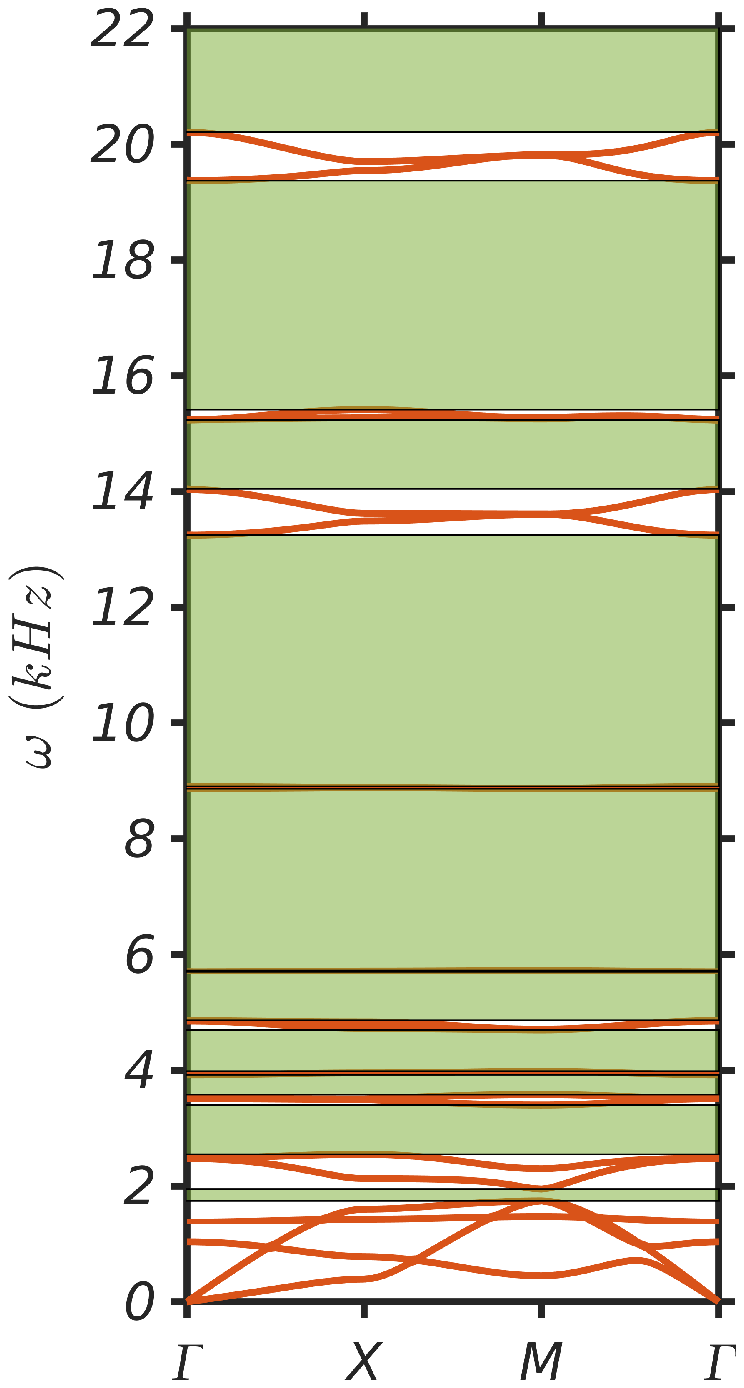}}
	\subfloat[{}]{\includegraphics[width=4cm,height=8cm]{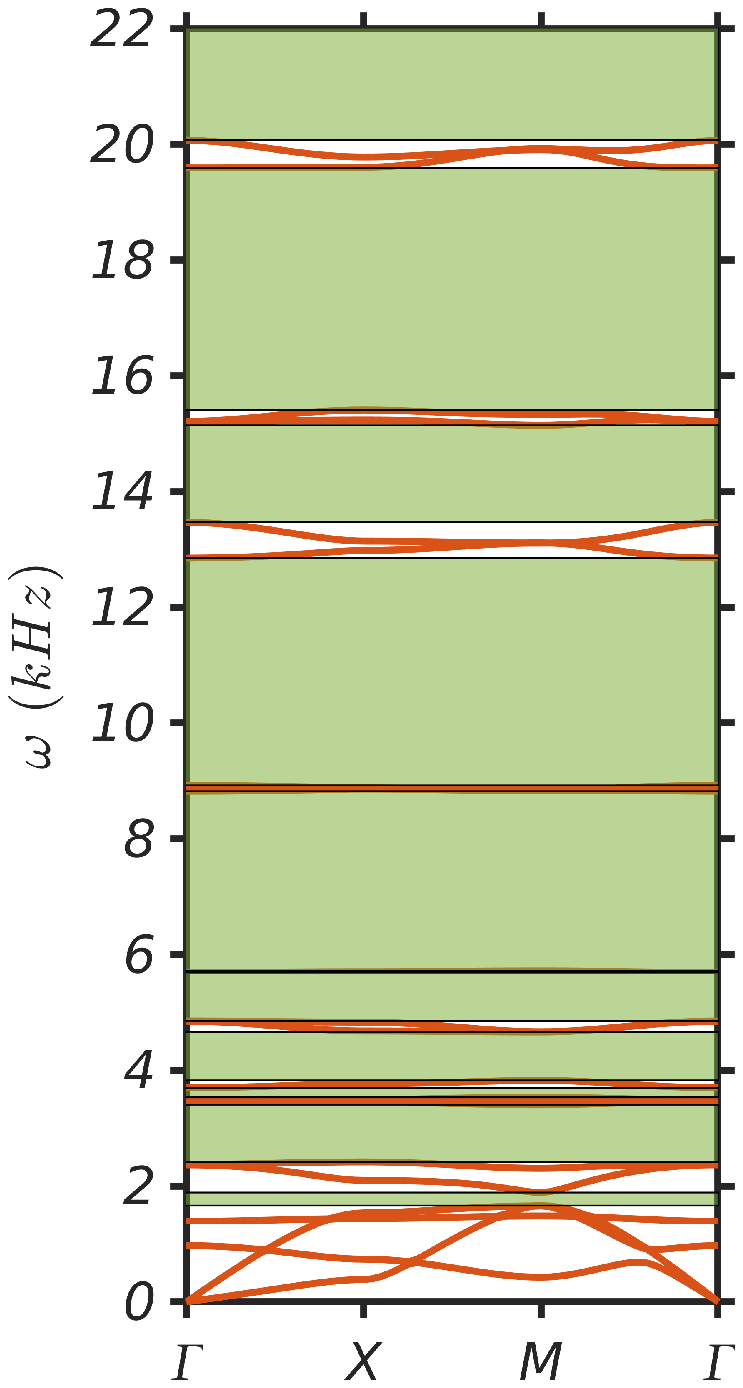}}\\
	\subfloat[{}]{\includegraphics[width=4cm,height=8cm]{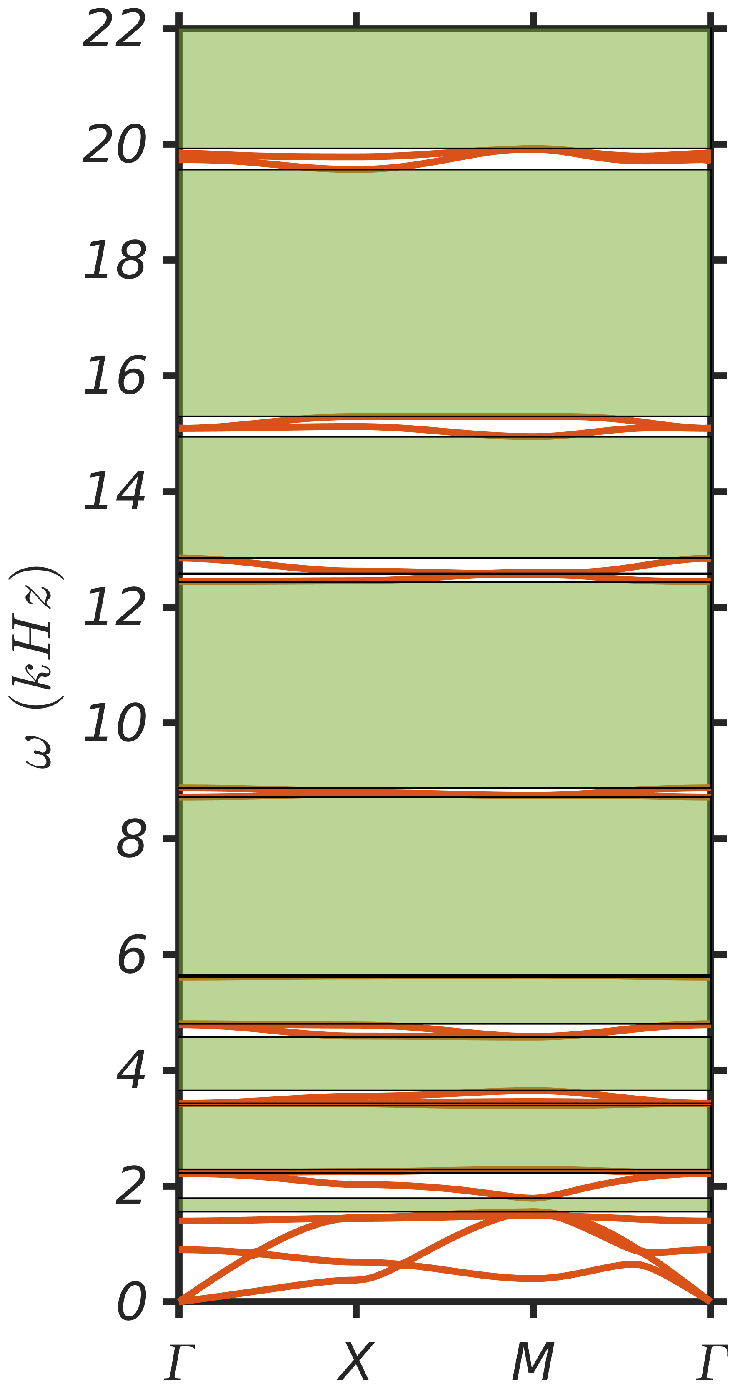}}
	\subfloat[{}]{\includegraphics[width=4cm,height=8cm]{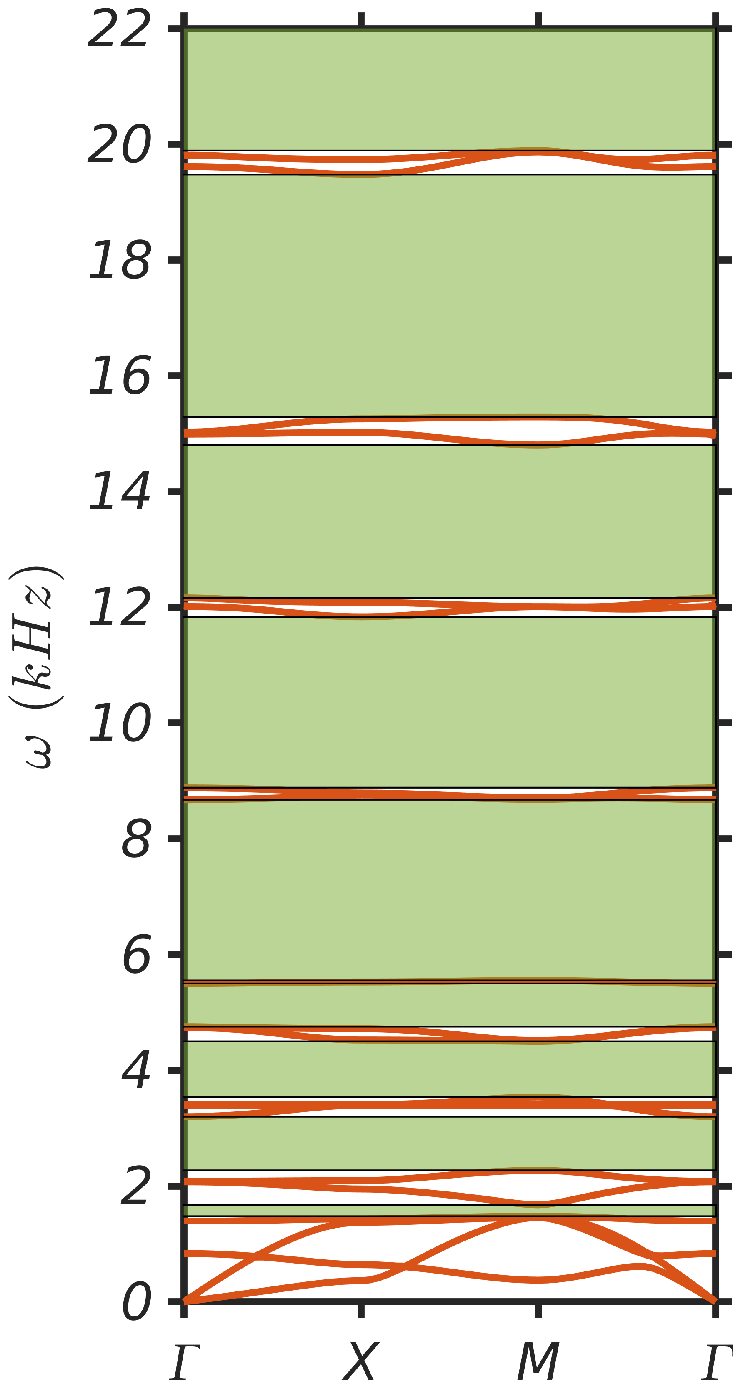}}
	\subfloat[{}]{\includegraphics[width=4cm,height=8cm]{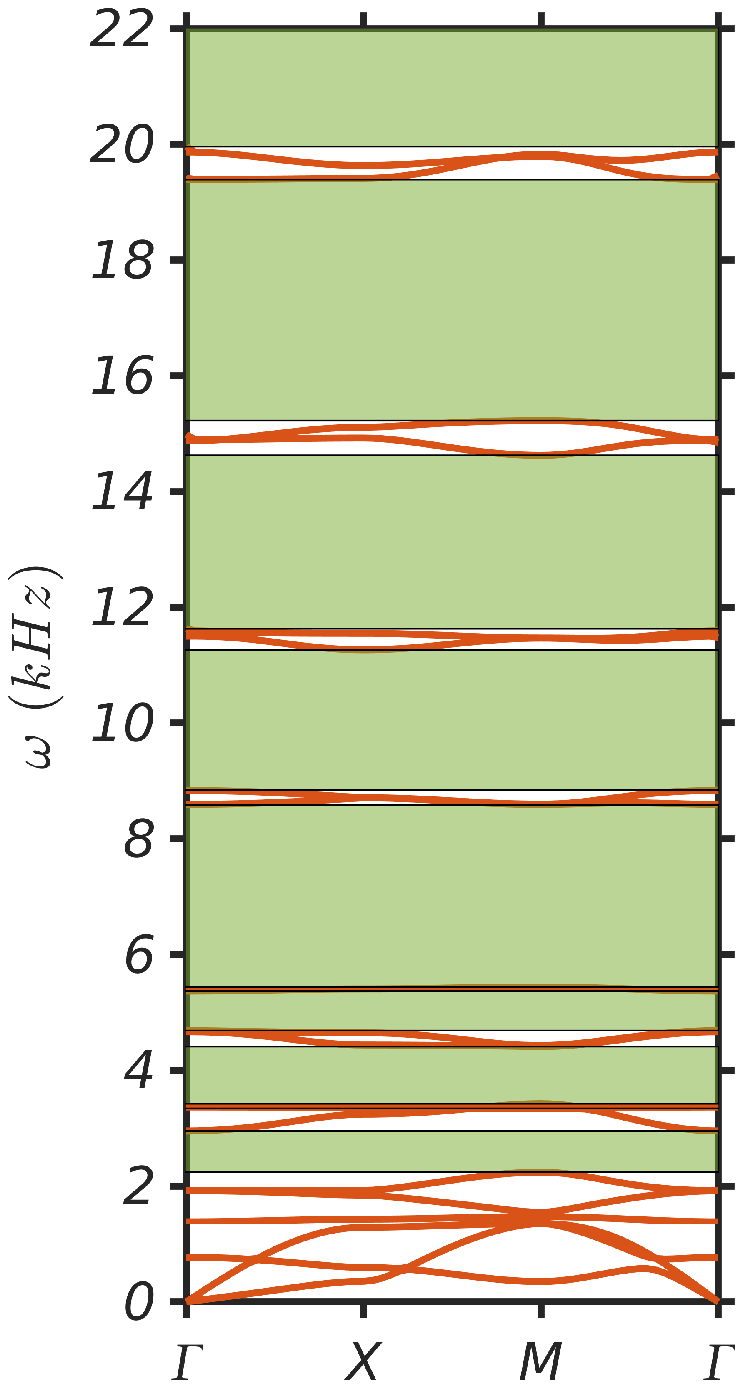}}
	\caption{Dispersion curves for diameter of localized mass (a) $D_{m}$=$3\,mm$, (b) $D_{m}$=$3.5\,mm$, (c) $D_{m}$=$4\,mm$, (d) $D_{m}$=$4.5\,mm$, (e) $D_{m}$=$5\,mm$, (f) $D_{m}$=$5.5\,mm$, and (g) $D_{m}$=$6\,mm$ along $\Gamma$-$X$-$M$-$\Gamma$ path.}
	\label{fig:Bandgapresultdiameter}	
\end{figure*}

\begin{figure}[!htb]  
	\centering
	\subfloat{\includegraphics[width=0.4\textwidth]{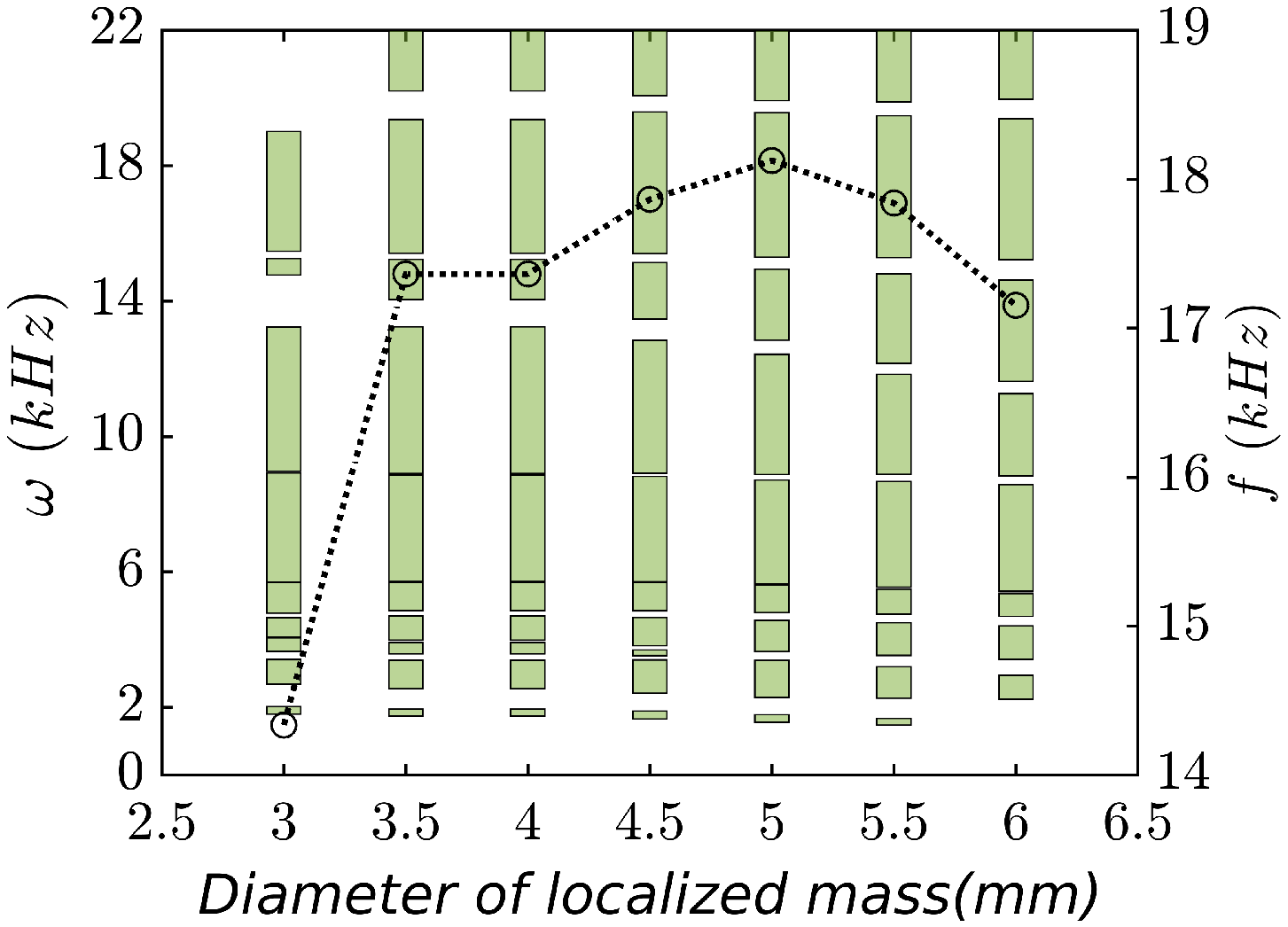}}
	\caption{Bandgaps and effective bandgap width vs diameter of localized mass ($D_{m}$) with $L_{2}$= $6\,mm$, and $\theta$=$30^o$.}
	\label{fig:Effbandgapdiameter}	
\end{figure}

Further, by changing the diameter of the localized mass connected to the struts dispersion responses are simulated and displayed in Fig.~\ref{fig:Bandgapresultdiameter}. It is marked from Fig.~\ref{fig:Bandgapresultdiameter} that bandgaps become wide and uninterrupted with the accession of the diameter of localized mass. Particularly, variations of bandgaps are very insensitive with the increment of the diameter of localized mass after $D_{m}$= $3.5\,mm$. Moreover, variation of bandgaps and effective bandgap width with the diameter of localized mass is delineated in Fig.~\ref{fig:Effbandgapdiameter}. Interestingly, the effective bandgap width lies between 17 to 18 kHz  for $D_{m}$= $3.5$ to $6\,mm$. However, optimum bandgap width is obtained at $D_{m}$= $5\,mm$.  
Ultimately, by changing the length 
of struts ($L_2$), strut angle ($\theta$), and diameter of the localized mass the negative effective properties
can be tailored over various frequencies. 

\subsection{Variation of bandgaps and associated Bloch mode shapes for different strut angle ($\theta$)}

\begin{figure}[!htb]  
	\centering
	\setlength\abovecaptionskip{-0.6\baselineskip}
	\subfloat{\includegraphics[width=0.5\textwidth]{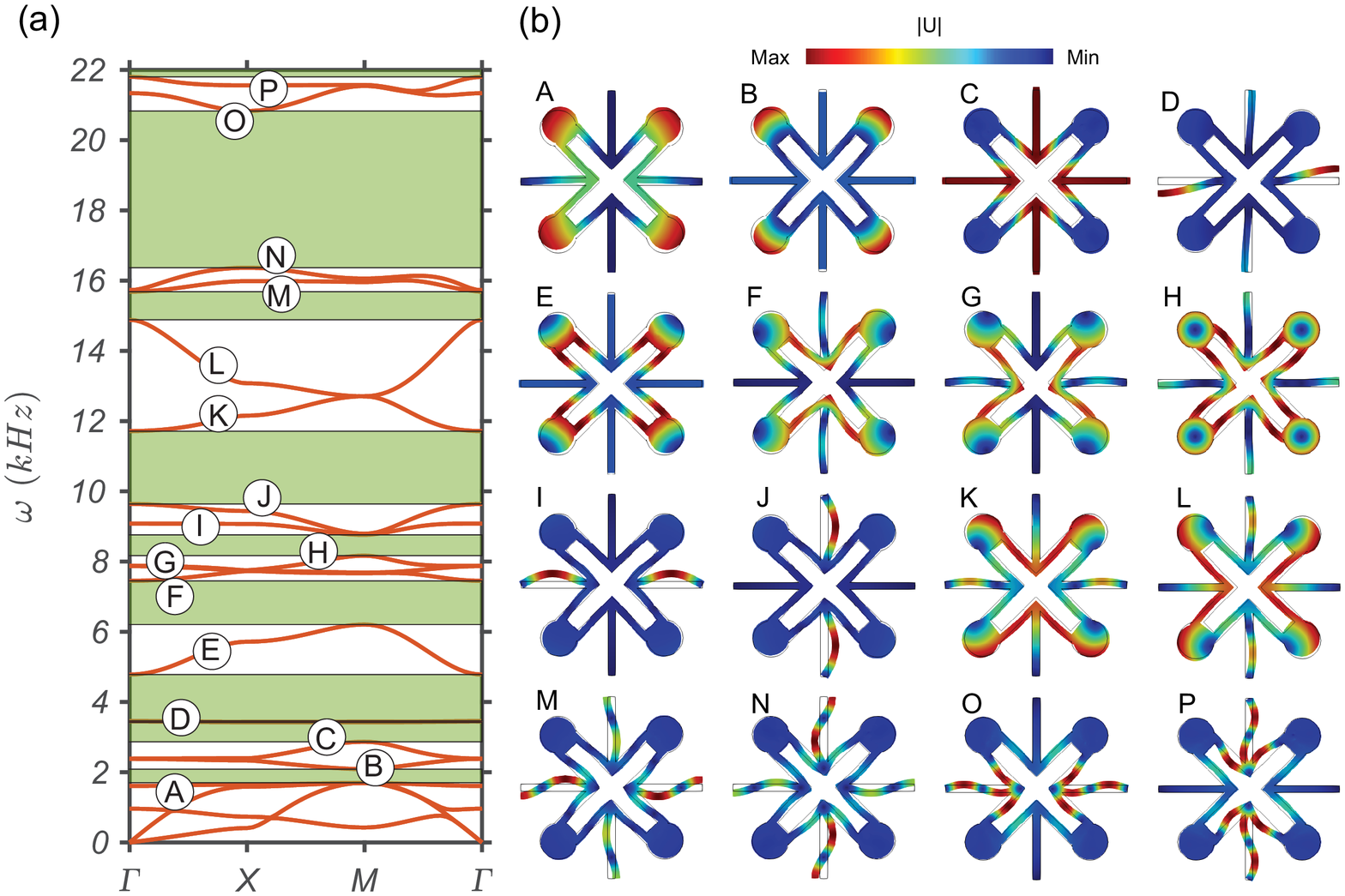}}
	\caption{(a) Dispersion curves of the metastructure with angle $\theta$=$45^o$ along  $\Gamma$-$X$-$M$-$\Gamma$ path, and (b) Bloch mode shapes at $M$ point for alphabetically marked bands.}
	\label{fig:BandgapresultSTARP45}	
\end{figure}

\begin{figure}[!htb]  
	\centering
	\setlength\abovecaptionskip{-0.6\baselineskip}
	\subfloat{\includegraphics[width=0.5\textwidth]{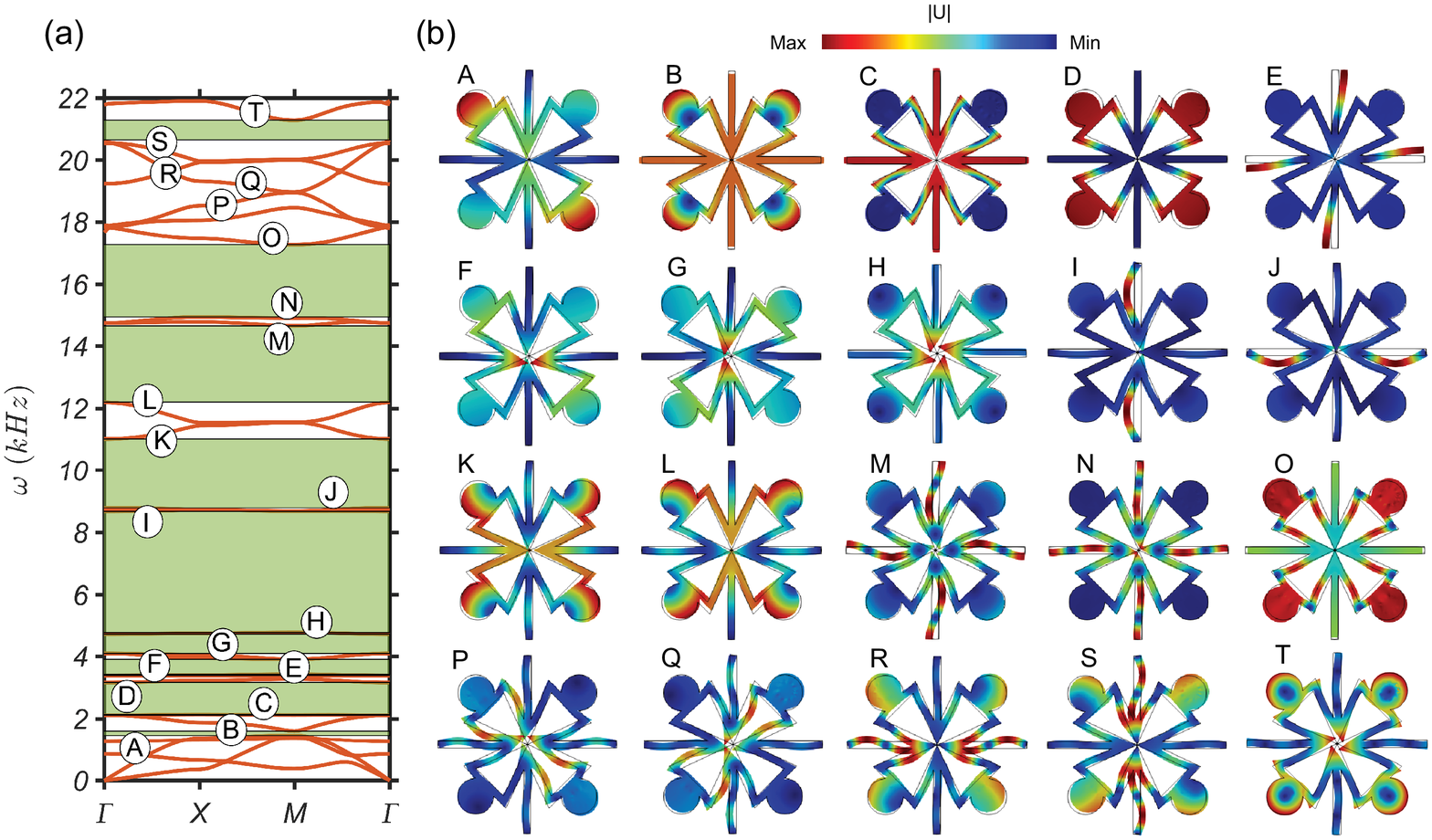}}
	\caption{(a) Dispersion curves of the metastructure with angle $\theta$=$25^o$ along  $\Gamma$-$X$-$M$-$\Gamma$ path, and (b) Bloch mode shapes at $M$ point for alphabetically marked bands.}
	\label{fig:BandgapresultSTARP25}	
\end{figure}

In the present study, the proposed single phase metamaterial exhibits the bandgap due to negative effective mass at dipole resonance~\citep{liu2000locally,yang2008membrane} and negative effective modulus at monopole/ quadrupole resonance~\citep{lai2011hybrid,jing2015soft}, respectively.
Dipole resonance originates through polarized oscillation of lumped mass at the inner junctions with lower order rotational symmetry~\citep{lai2011hybrid,jing2015soft,xia2019robust}. In contrast, non polarized oscillation of lumped mass at outer junctions
with two/four fold rotational symmetry can offer quadrupole/monopole resonance, respectively~\citep{lai2011hybrid,jing2015soft,li193multipolar}. 
The effective bandgap width remarkably changes mainly with strut angle as it varies from 18.12 kHz to 2.18 kHz as found in Fig.2(b) of the main article. 
Therefore,  
we primarily investigate the vibration modes to address the presence of dipole and monopole/quadrupole resonances at the 
frequency bands with various strut angles 
as indicated in Fig.~\ref{fig:BandgapresultSTARP25},~\ref{fig:BandgapresultSTARP45}, and Fig.~3 ($\theta$=$55^o$), Fig.~4 ($\theta$=$30^o$) of the main article. We especially
focus on the $M$ point ($\pi/a$, $\pi/a$) where the majority of the bandgaps are conceived
for the unit cell with numerous strut angles.
Bandgaps and its associated Bloch mode shapes for the unit cell with $\theta$=$45^o$ are presented in Fig.~\ref{fig:BandgapresultSTARP45}(a) and (b), respectively. The vibration modes are superimposed with undeformed  geometry outlined in black color.
Band A demonstrates the rotation of cross struts with localized mass in the opposite direction resulting in bending of horizontal struts with no deformation in vertical struts. This deformation mode manifests the dipole resonance. Meanwhile, band B imparts expansion and compression of horizontal and vertical struts and accommodates opposite rotation of cross struts. This attended deformation clearly marks the quadrupole resonances. Further, band C portrays the outward movement of horizontal and vertical struts consisting of expansion of cross struts and reveals monopole resonance.
Subsequently, the band D delineates the asymmetric magnitude of the bending deformation of horizontal and vertical struts, while the core remains stationary. Hence, this signature again denotes the quadrupole resonance.
In succession, the band E signifies
the bending deformation of cross struts with localized mass subjected to the compression/extension of straight struts evolving quadrupole resonance. Afterwards, bands F and band G show identical deformation with dissimilar orientations. Consequently, these vibration modes contribute the same effective properties resulting in no bandgap at $M$ point~\citep{wang2004two,wang2014harnessing}.
In parallel, the band H exemplifies the rotation of cross struts which causes the slight bending of straight struts noted as the monopole resonance. 
Eventually,
Bloch modes of bands I and J imply the deformation of two straight struts while
other struts stand as undeformed. These modes are akin to the
Bloch modes of conventional LRAMs appearing in dispersion corresponding to the
standing waves. This behavior is interpreted as the emergence of a stationary core while the soft coating is subjected to
rotation under dynamic conditions~\citep{krushynska2014towards,kumar2019low}. 
Furthermore, no bandgap is achieved between bands K and L, bands M and N, as well as for bands O and P at $M$ point.
We can emphasize that a unit cell with localized masses on the cross struts has fewer bands that have similar Bloch modes with different orientations as contrasted to the struts with $55^o$(Fig. 3 of the main article). Because of this, bandgaps evolve in the intermediate frequency range for angle $\theta$=$45^o$ as compared to the angle $\theta$=$55^o$.

Moreover, the high magnitude of deformation is noticed for quadrupole and monopole resonance at angle $\theta$=$30^o$ for band B and C (Fig. 4 of the main article) as compared to the analogous bands of cross struts $\theta$=$45^o$ as shown in Fig.~\ref{fig:BandgapresultSTARP45}. Consequently, fewer bands are observed with similar Bloch modes with diverse orientations at $M$ point resulting in larger bandgaps for the metastructure with $\theta$=$30^o$.
Afterwards, effective bandgap width decreases at angle $\theta$=$25^o$ as presented in Fig.~2(b) of the main article. In order to understand the transition of this bandgap formation, associated Bloch mode shapes for the unit cell with $\theta$=$25^o$ are displayed in Fig.~\ref{fig:BandgapresultSTARP25}. Analogous to the previous cases, bands A, B, and C represent the dipole, monopole, and quadrupole resonance, respectively.
Besides, the band D interprets the rotation of struts with localized mass in different directions while horizontal and vertical struts prevail static. Hence, the quadrupole resonance of the unit cell becomes evident.
Thereafter, the band E declares the bending of four straight struts in the opposite direction while the core remains stationary which confirms again the presence of quadrupole resonance.
In succession, bands F and G show identical deformation modes but in a dissimilar orientation such that no bandgap is build up at the $M$ point. Next, the band H exhibits the twisting of inner junctions which causes the slight deformation of horizontal and vertical struts marked as the torsional mode. In this mode, deformation is limited to the core and hence no force is exerted to the structure by the core.
Accordingly, due to the lack of interaction of the core with the external wave field, no bandgap is evolved at $M$ point~\citep{wang2004two,krushynska2014towards}.
Later, bands I and J manifest the deformation of two straight struts while
other struts stand as straight resulting in the standing waves. 
Similarly, bands K and band L denote similar deformation modes subjected to disparate directions. 
Meanwhile, band M delineates the bending of the diagonal struts with localized mass subjected to the deformation of straight struts dictates the monopole resonance. Likewise, band N indicates the inward and outward deformation of diagonal struts with localized mass and bending of four straight struts which convey the quadrupole resonance. 
Consequently, the band O suggests the movement of straight struts in an inward direction while the motion of localized mass occurs in an outward direction. Due to the four fold rotational symmetry of this mode, this behavior clearly reports the monopole resonance. 
In contrast, the bands P and Q are occasioned by outward and inward movement of one side of diagonal struts with the deformation of straight struts. These two modes are comparable in different directions. As well as, bands R and S illustrate similar deformation modes with diverse orientations. Subsequently, band T portrays the four fold rotational symmetry and marks the monopolar resonance.
Thus, identical deformation with distinctly oriented quadrupoles is noted between bands P and Q, as well as for bands R and S in the higher frequency range at angle $\theta$=$25^o$. 
As a result, higher frequency bandgap vanishes, and overall effective bandgap width decreases as compared to angle $\theta$=$30^o$. 

\begin{figure}[!htb]  
	\centering
	\includegraphics[width=0.48\textwidth]{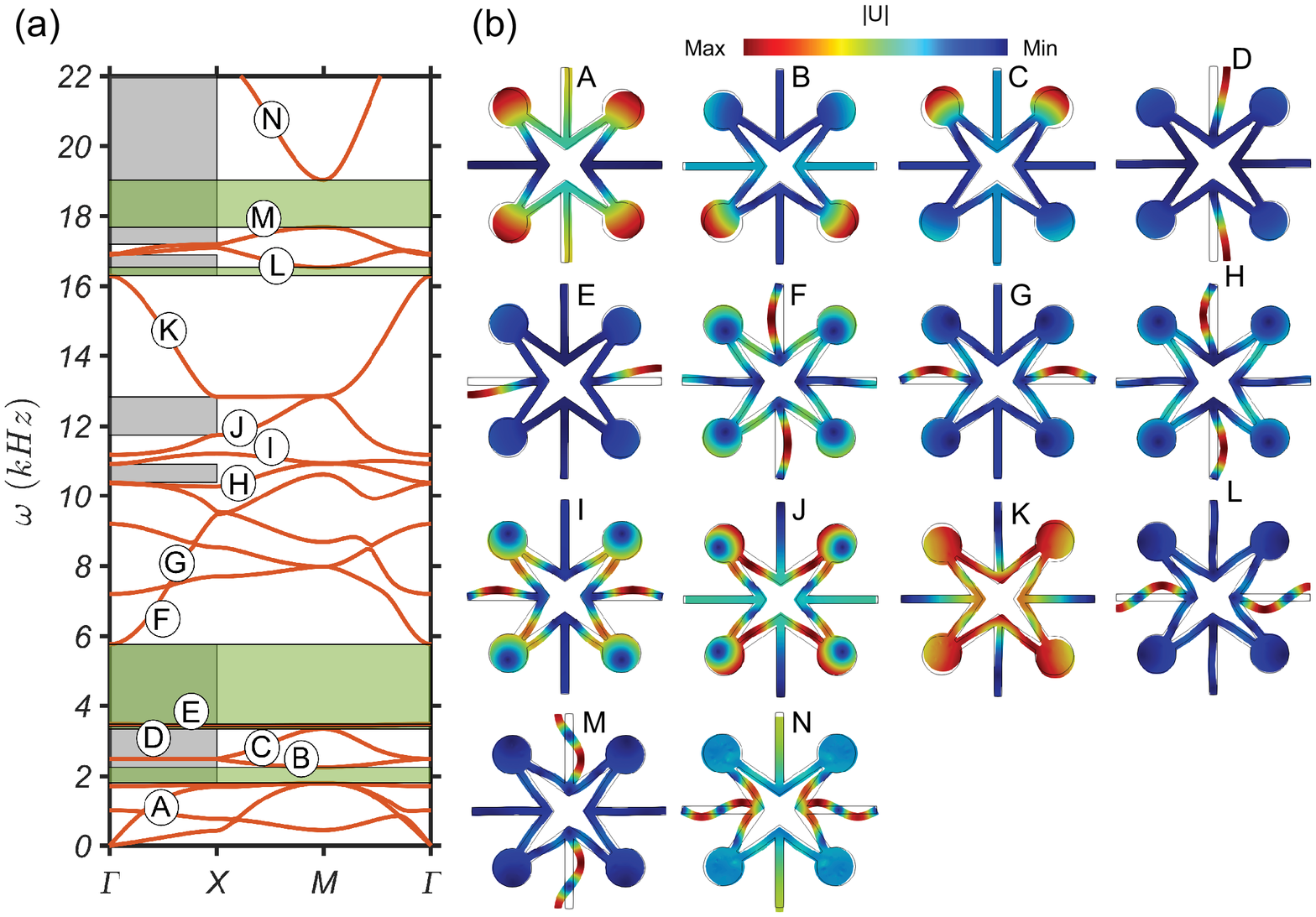}
	\caption{(a) Dispersion curves of the metastructure for $\theta$=$55^o$ where, green and gray color zones indicate the
		omnidirectional and directional bandgaps, respectively, and (b) Bloch mode shapes at $X$ point for alphabetically marked bands. Color bar represents the magnitude of absolute displacement.}
	\label{fig:DirectionalModeshapes}
\end{figure}

Further, to understand the directional bandgap formation of the metastructure with angle $\theta$=$55^o$, we have examined the Bloch mode shapes at $X$ point as displayed in Fig.~\ref{fig:DirectionalModeshapes}(b).
In this metastructure directional bandgaps are evident from 1.8 to 5.8 kHz, 10.4 to 10.9 kHz, 11.75 to 12.8 kHz, 16.3 to 16.9 kHz, and 17.2
to 22 kHz as marked by the shaded gray zone as presented in Fig.~\ref{fig:DirectionalModeshapes}(a). Directional bandgaps are also analyzed numerically and their existence has been validated experimentally by several researchers~\citep{zhang2021realization,sun2019band,tian2020perforation}.
In comparison to the Bloch mode shapes at $M$ point (Fig.~3(b) of the main article), quadrupole resonance evolves for the bands H, I, J, and K at $X$ point. Consequently, multiple bandgaps are generated in the intermediate frequency range at $X$ point. Moreover, vibration modes of similar deformation with different directions are noticed between bands B and C, bands D and E, as well as for bands L and M leading no bandgap at $X$ point. In parallel, high amplitude of deformation is noted for dipolar resonance (at band A) and bending dominated quadrupole resonance (at band N) at $X$ point. Therefore, owing to higher negative effective properties, a large frequency bandgap is developed  in lower and higher frequency regimes at $X$ point as compared to the $M$ point.

\section{Experimental Procedure}
\begin{figure}[!htb]  
	\centering
	\subfloat[{}]{\includegraphics[width=0.23\textwidth]{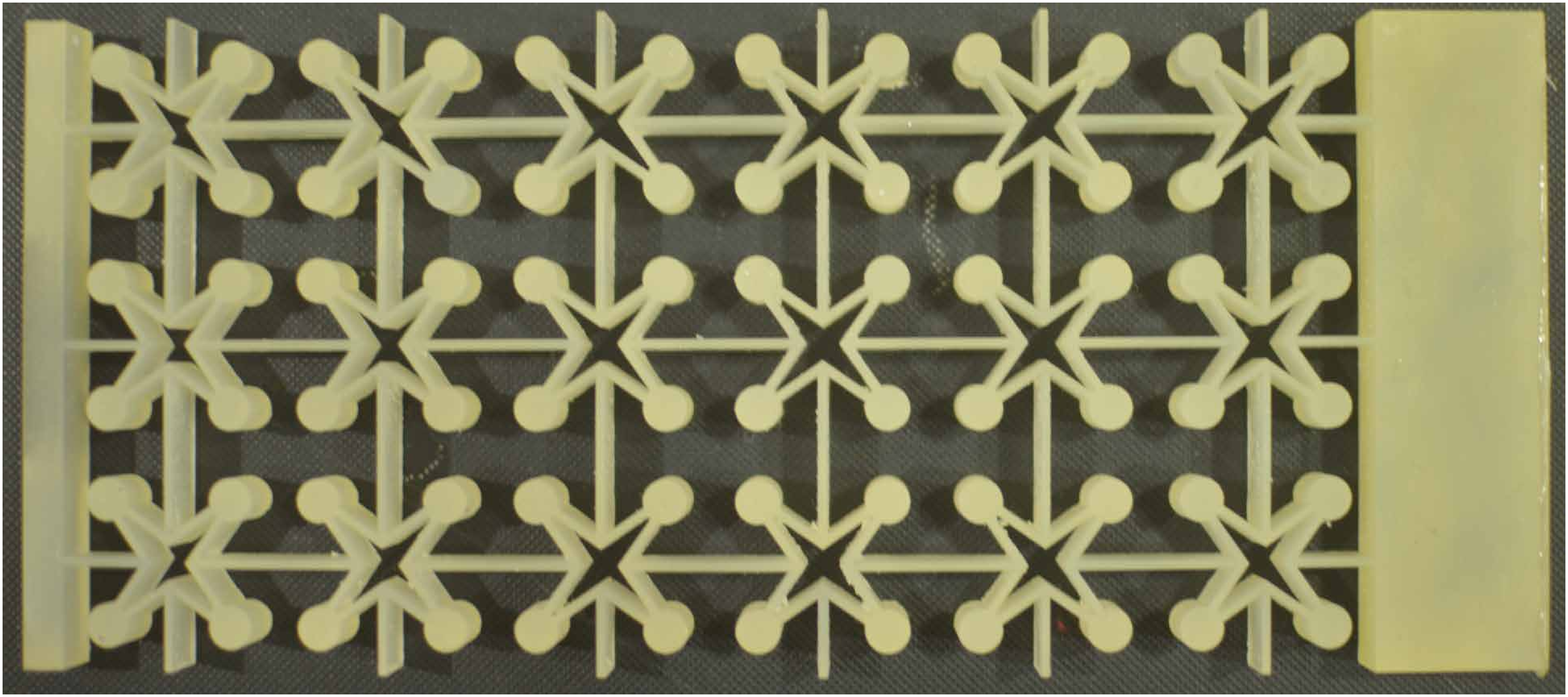}} \hspace{0.2 cm} 
	\subfloat[{}]{\includegraphics[width=0.23\textwidth]{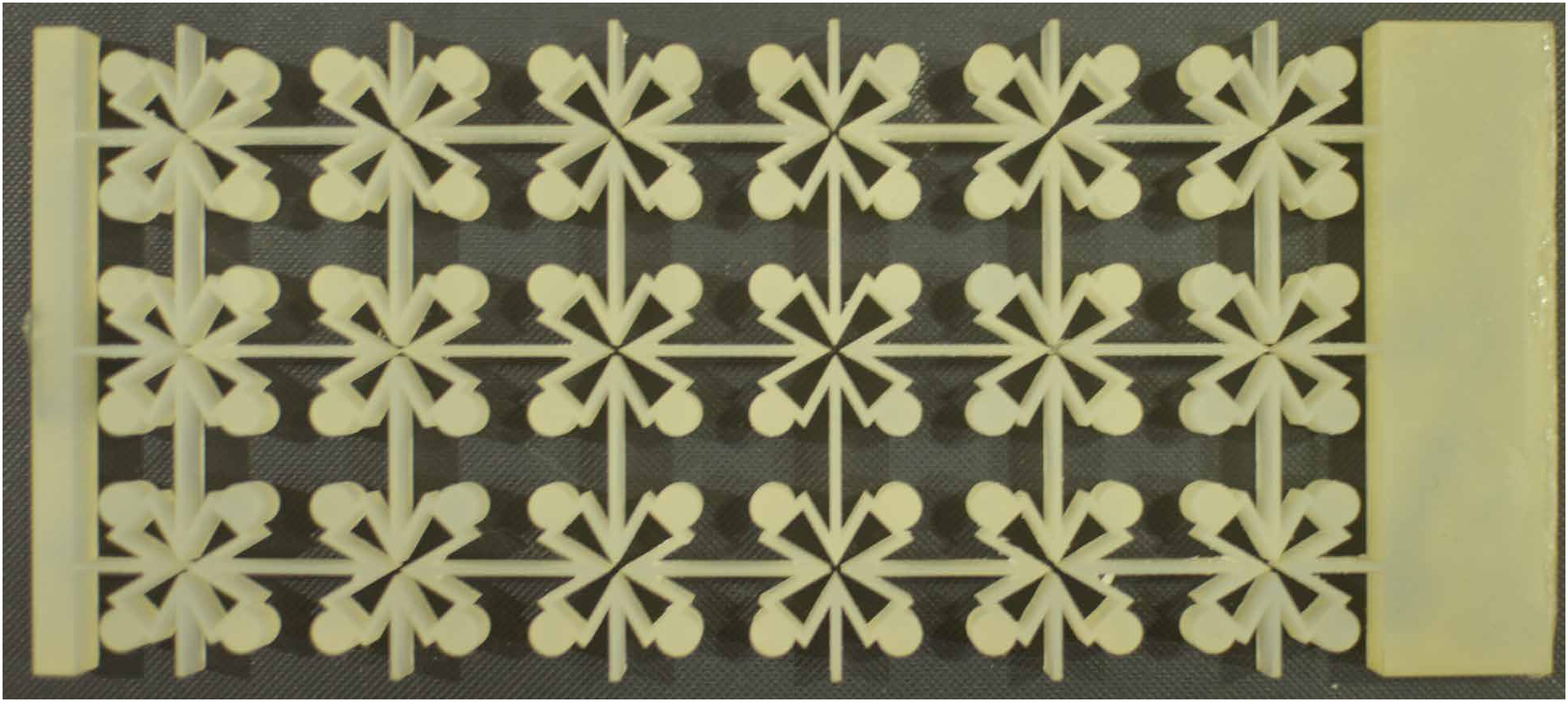}}
	\caption{3D printed metamaterials with strut angle (a) $\theta$=$55^o$, and (b) $\theta$=$30^o$.}
	\label{fig:specimen}	
\end{figure}

In order to validate the numerically predicted bandgap, an experiment of mechanical wave transmission is conducted on the finite size of fabricated metamaterial (length $L$=150 $\,mm$, width $W$=75 $\,mm$, and 
thickness $H$=10 $\,mm$). Specimens are fabricated using Visijet M3 crystal from 3D printers (Projet MJP 3600, 3D Systems) in periodically arranged
multiple unit cells (3$\times$6 matrix)~\citep{bilal2018architected,yao2008experimental,kumar2019low} with cylinder of same material connected to the struts with angle $\theta$=$55^o$ and $\theta$=$30^o$ as delineated in Fig.~\ref{fig:specimen}(a) and (b), respectively.  
Sine wave signals of various frequencies are generated with the Data Acquisition System (VibSoft5.3, Polytec Inc.) through an in-built function generator. 

%
%
\begin{figure}[!htb]  
	\centering
	\includegraphics[width=0.5\textwidth]{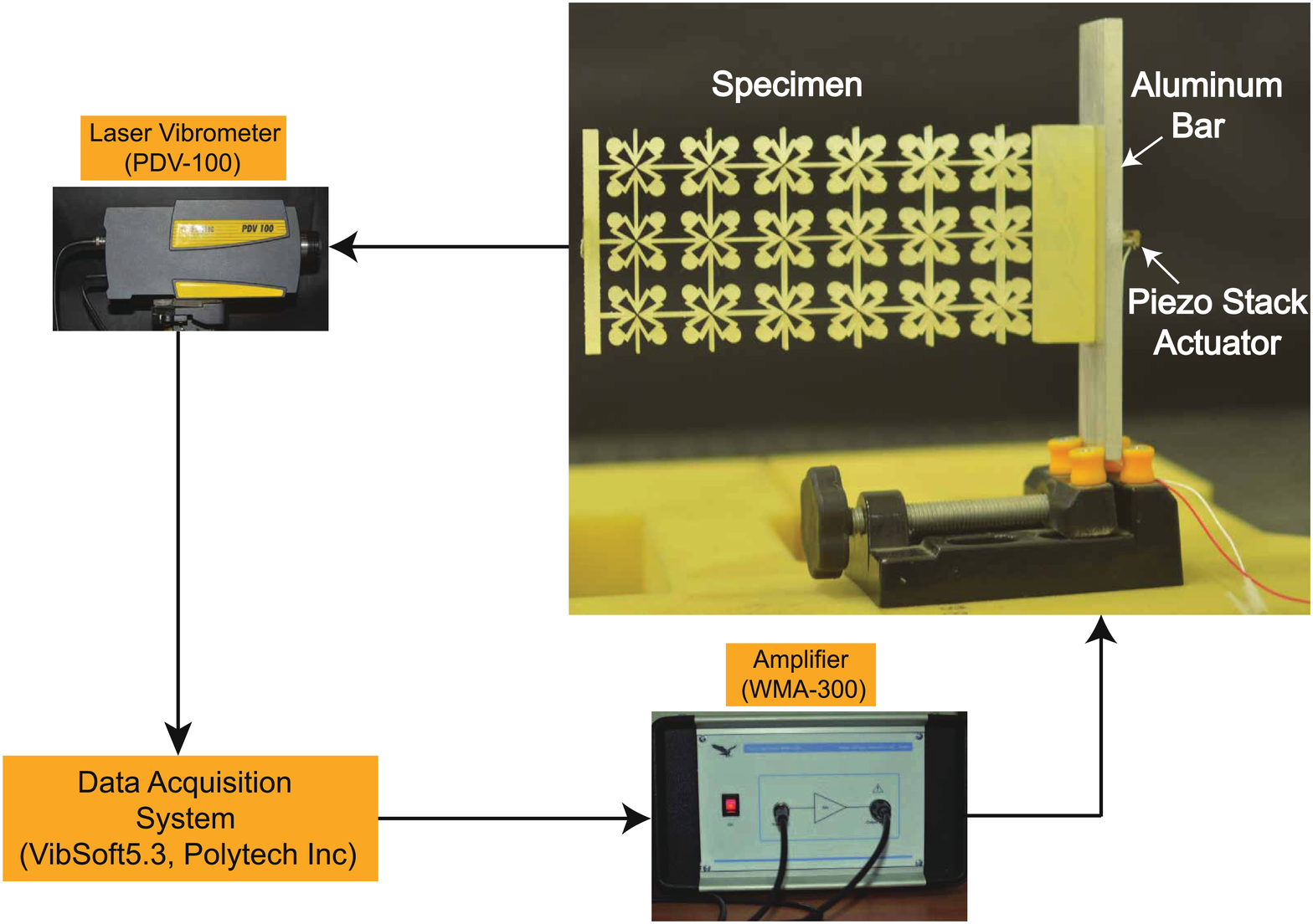}
	\caption{Flow chart for experiment of mechanical wave generation and data acquisition for extracting transmission spectrum at the free end of specimen.}
	\label{fig:flowchart}
\end{figure}
An experimental flow chart for extracting the transmission spectrum is shown in  Fig.~\ref{fig:flowchart}. Distinct frequencies of sine wave signals are produced with a function generator. At first, an electrical signal is amplified through a high voltage amplifier (WMA-300, Falco Systems), then converted to a mechanical signal by a piezo stack actuator (NAC2013-H10, Noliac Inc.). Moreover, the piezo stack actuator adheres with an aluminum bar which is mounted to the fabricated specimen for plane wave propagation. Afterwards, the assembled specimen is placed horizontally~\citep{oh2018zero,huang2017periodic,yao2008experimental} through a small mechanical vice and kept on the soft pad (Fig.~\ref{fig:flowchart}). 
Utilizing, portable Laser Doppler Vibrometer (PDV-100, Polytec Inc.), the incident mechanical signal (velocity signal) is extracted on the aluminum bar. Further, the output velocity signal is collected on the left end of the specimen.
The incident and transmitted velocity signals are obtained in 50 Hz to 22 kHz by sweeping the frequency from 50 Hz to 22 kHz over 0.02 seconds with an interval of 50 Hz.
We take the sampling frequency, $f_s$=102.4 kHz, and the number of FFT lines is 6400 resulting in a frequency resolution of
6.25 Hz. 
Finally, a transmission spectrum is evaluated by converting the time domain signal into the frequency domain using Fast Fourier Transformation (FFT) as presented in Fig. 5 of the main article. We demonstrate the omnidirectional and directional bandgaps of the metastructure for angle $\theta$=$55^o$ in Fig. 5(a) of the main article, while we discuss the omnidirectional bandgaps for the metastructure with $\theta$=$30^o$ in Fig. 5(b).
Experimental data of the input and output velocity signals for the metastructures with angle $\theta$=$55^o$, and $\theta$=$30^o$ are provided in supplemental data sheets 1, and 2, respectively. We further carried out the experiment to analyze the transmission with sine signals of fixed frequency by selecting the bandpass filter as delineated in Fig. 6 of the main article. In addition, we assume that the transmission spectra are merely contributed from the material damping due to it having negligible loss factor under ambient conditions~\citep{kumar2019low}.

\end{document}